\documentclass[twocolumn,prl,nobalancelastpage,aps,10pt,nofootinbib, notoc]{revtex4-1}

\usepackage{graphicx,bm,times}
\usepackage{amsmath}
\usepackage{float}
\usepackage{subfig}
\usepackage{blindtext}
\usepackage{enumitem}
\usepackage[english]{babel}
\usepackage[autostyle]{csquotes}
\usepackage{adjustbox}
\usepackage{xcolor}
\usepackage[bottom]{footmisc}
\usepackage[font={small,it},justification=RaggedRight]{caption}
\usepackage{textcomp}
\usepackage{mathtools}

\DeclarePairedDelimiter\ket{\lvert}{\rangle}
\DeclarePairedDelimiterX\braket[2]{\langle}{\rangle}{#1 \delimsize\vert #2}
\usepackage{xfrac}
\usepackage{array}
\usepackage{gensymb} 

\usepackage{color}   
\usepackage{hyperref}
\hypersetup{
    colorlinks=true, 
    linktoc=all,     
    linkcolor=blue,
    citecolor=blue,
}
\hypersetup{linktocpage}

\renewcommand{\thesection}{\arabic{section}}
\renewcommand{\thesubsection}{\thesection.\arabic{subsection}}
\renewcommand{\thesubsubsection}{\thesubsection.\arabic{subsubsection}}

\usepackage{titlesec}
 
\titleformat{\section} 
	{\normalfont\Large\bfseries}{\makebox[20pt][l]{\thesection}}{0pt}{} 
\titleformat{\subsection} 
	{\normalfont\large}{\makebox[20pt][l]{\thesubsection}}{0pt}{}
\titleformat{\subsubsection} 
	{\normalfont\normalsize}{\makebox[20pt][l]{\thesubsubsection}}{4pt}{}

\makeatletter
\newcommand*{\balancecolsandclearpage}{%
  \close@column@grid
  \cleardoublepage
  \twocolumngrid
}
\makeatother

\begin{document}
\begin{titlepage}

\newcommand{\HRule}{\rule{\linewidth}{0.5mm}} 

\center 
 
\includegraphics[width=55mm]{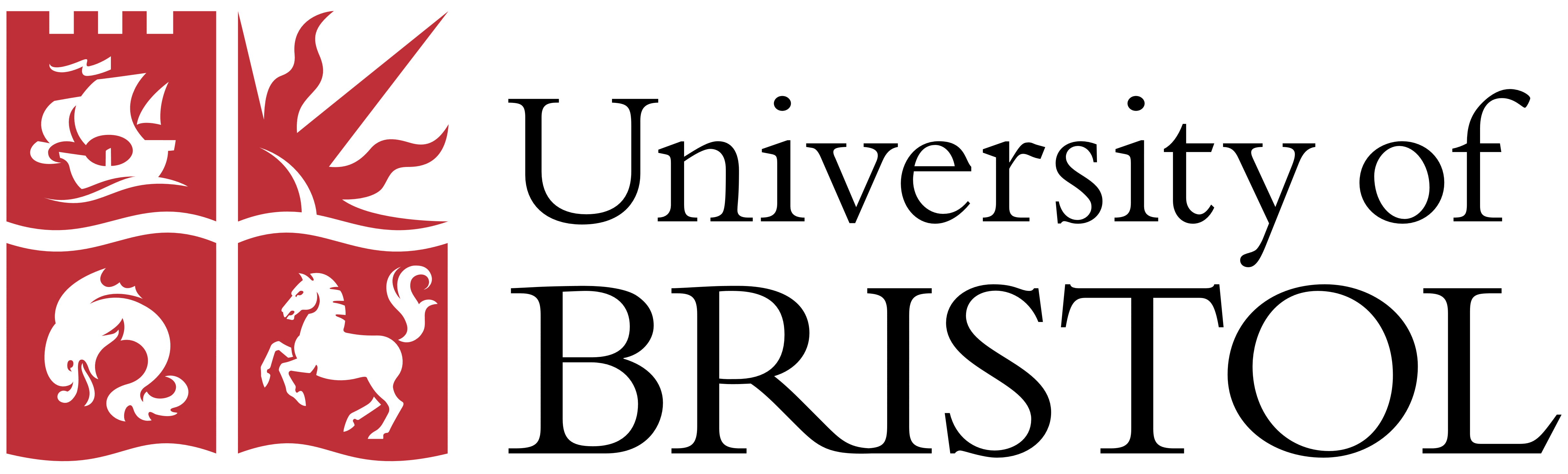}\\[0.5cm] 
\vspace{+0.7cm}
\textsc{\Large Department of Physics}\\[0.4cm] 
\textsc{{May 2021}}\\[0.4cm] 


\HRule \\[0.6cm]
{ \huge Future of computing at the Large Hadron Collider  
 }\\[0.4cm] 
\HRule \\[1.8cm]


\begin{minipage}{0.4\textwidth}
\begin{flushleft} \large
\emph{Author:}\\
Dhananjay Saikumar 
\end{flushleft}
\end{minipage}
~
\begin{minipage}{0.4\textwidth}
\begin{flushright} \large
\emph{Academic Supervisors:} \\
Dr Daniel O\textquotesingle Hanlon 
\& 

Prof Jonas Rademacker 

\end{flushright}
\end{minipage}\\[1.4cm]
\vspace{+0.2cm}
\includegraphics[width=50mm]{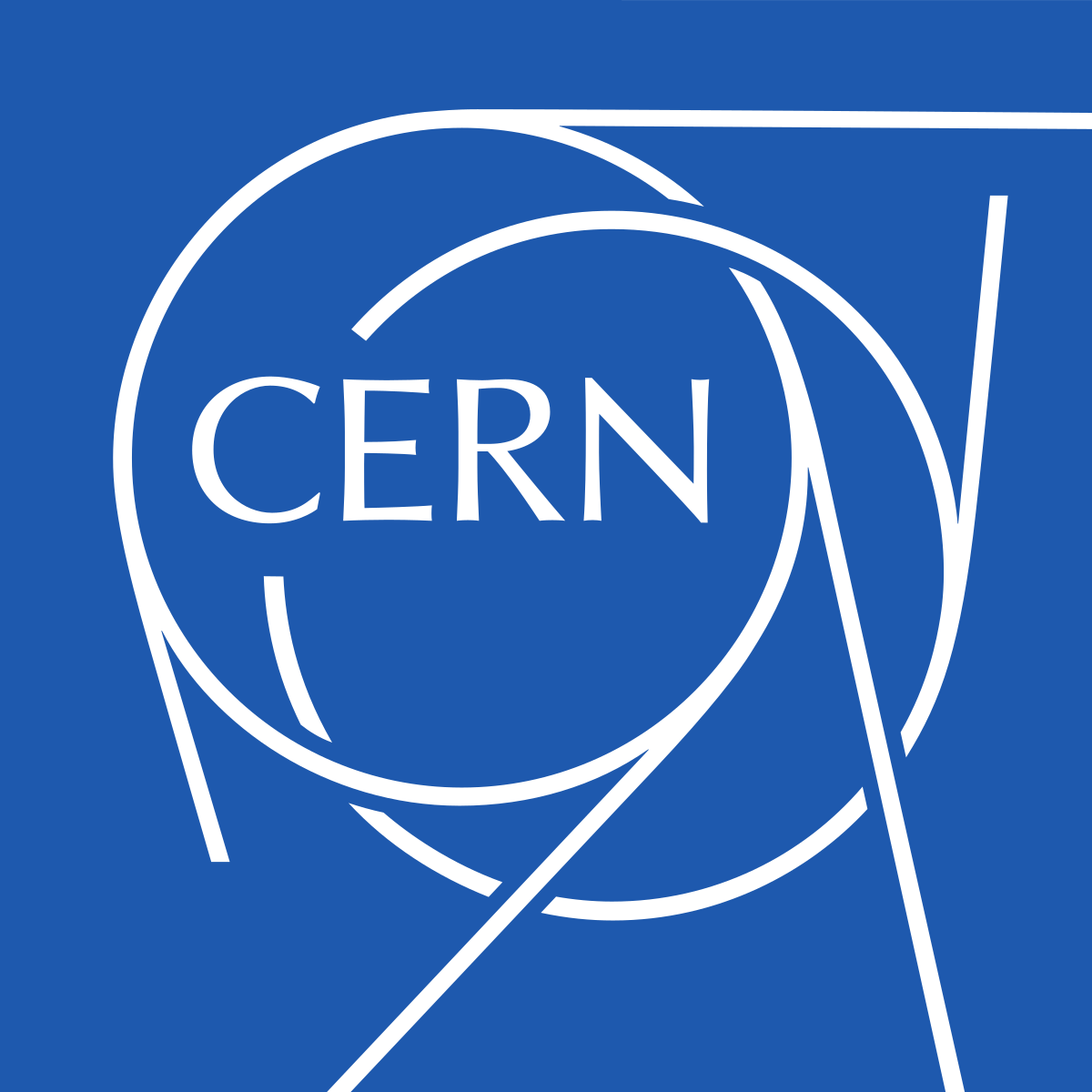}\\[0.5cm] 
\textsc{ {For submission in 2021}}\\[0.4cm] 
\vspace{+3cm}
\end{titlepage}
\onecolumngrid
\clearpage
\newpage

\newpage
\section*{Abstract}
High energy physics (HEP) experiments at the LHC generate data at a rate of $\mathcal{O}(10)$ Terabits per second. This data rate is expected to exponentially increase as experiments will be upgraded in the future to achieve higher collision energies. The increasing size of particle physics datasets combined with the plateauing single-core CPU performance is expected to create a four-fold shortage in computing power by 2030. This makes it necessary to investigate alternate computing architectures to cope with the next generation of HEP experiments. This study provides an overview of different computing techniques used in the LHCb experiment (trigger, track reconstruction, vertex reconstruction, particle identification).
Furthermore, this research led to the creation of three event reconstruction algorithms for the LHCb experiment. These algorithms are benchmarked on various computing architectures such as the CPU, GPU, and a new type of processor called the IPU, each roughly containing $\mathcal{O}(10)$, $\mathcal{O}(1000)$, and $\mathcal{O}(1000)$ cores respectively. This research indicates that multi-core architectures such as GPUs and IPUs are better suited for computationally intensive tasks within HEP experiments.
    
\newpage
\onecolumngrid
	\tableofcontents
\newpage
\twocolumngrid

\section{Introduction}
\begin{figure}[H]
\includegraphics*[width=1\linewidth,clip]{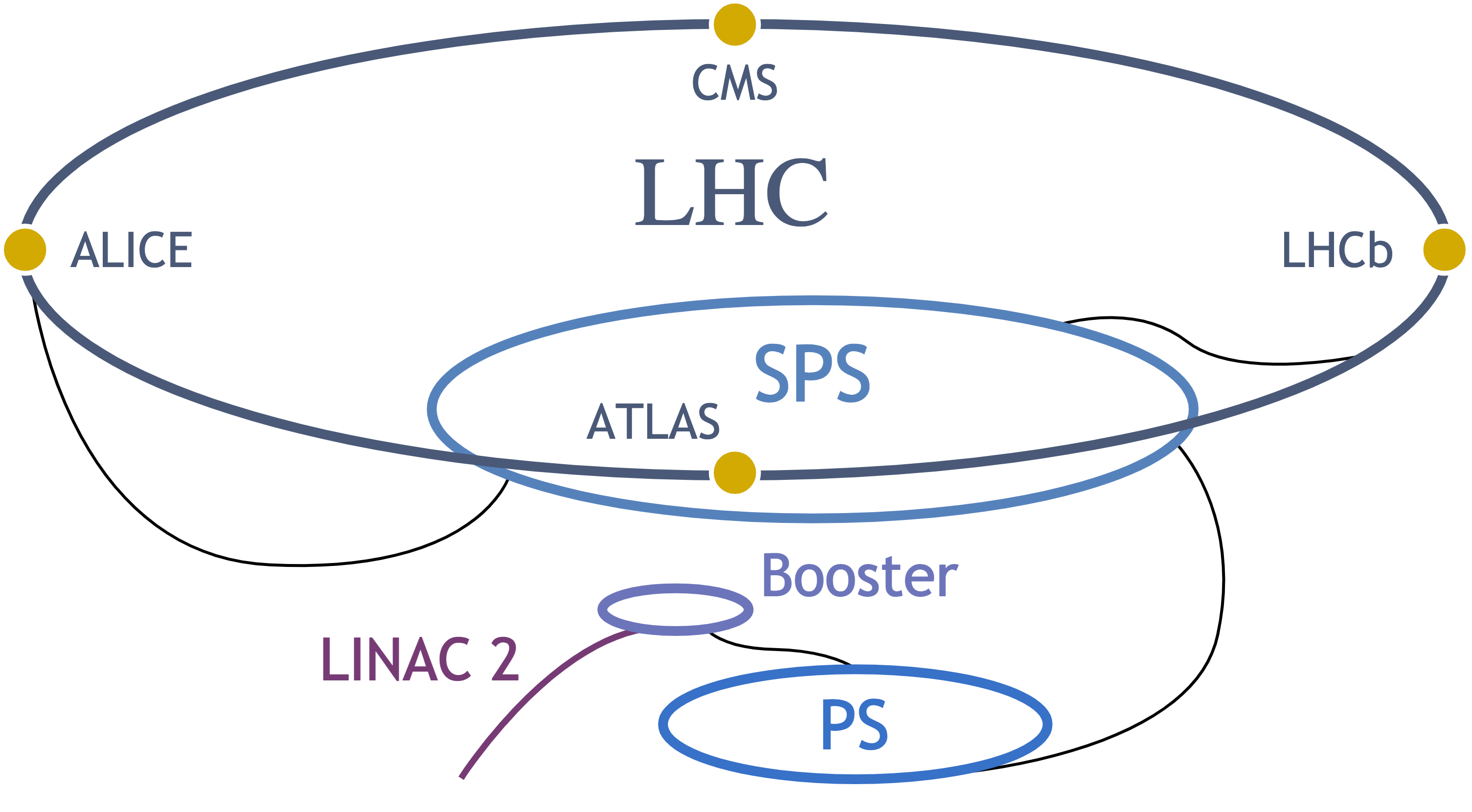}
\caption{The top of the figure represents the LHC and positions of the four major experiments. The bottom of the figure depicts small accelerators which inject hadrons into the 27KM LHC ring. Figure taken from Ref \cite{6}.} \label{fig:1}
\end{figure}

The Large Hadron Collider (LHC), located in Geneva Switzerland, is the largest particle accelerator ever built to-date, designed and constructed by the European Organization for Nuclear Research (CERN). The LHC is made up of a 27-kilometre underground tunnel filled with strong magnets that guide and boost beams of protons up-to velocities 3.1 m/s slower than the speed of light \cite{1}. Protons beams travelling in the opposite direction are designed to collide in four distinct regions corresponding to the four major particle physics experiments at the LHC (see figure \ref{fig:1}):

\begin{itemize}[leftmargin=+0.4cm]
\itemsep-0.27em 
\item[--] A Toroidal LHC Apparatus (ATLAS): Largest General purpose experiment at the LHC, designed to investigate a wide spectrum of physics, ranging from: the Higgs boson, CP violation\footnote{Charge conjugation Parity symmetry}, BSM\footnote{Physics beyond the Standard Model}, Dark matter, and more \cite{2}.

\item[--] Compact Muon Solenoid (CMS): Another General purpose experiment at the LHC which works in conjunction with the ATLAS Experiment (\textbf{e.g}: the joint discovery of the Higgs boson in 2012). It has the same scientific goals as ATLAS; however, the CMS detector operates on a completely different design approach \cite{3}. 

\item[--] Large Hadron Collider beauty (LHCb): The LHCb is a specialised experiment designed to study the effects of CP violation in beauty-hadron systems (shedding light on the matter anti-matter asymmetry observed in the universe), measuring forward-backward asymmetry in FCNC\footnote{Flavor-changing neutral currents} decays, searching for BSM phenomena, and more \cite{4}.

\item[--] A Large Ion Collider Experiment (ALICE): ALICE is designed to study heavy-ion physics. Collisions of Pb-Pb nuclei allow ALICE to investigate the fifth state of matter: the quark-gluon plasma (QGP). In this state, quarks and gluons are disentangled, resulting in conditions similar to a fraction of a second after the Big bang \cite{5}. 
\end{itemize}

\setlength\belowcaptionskip{-2ex}
The energy of the colliding particle beams is transformed into matter via Einstein's mass-–energy relation eq(1) in the COM\footnote{Center of momentum} frame, creating a wide shower of particles. 
\begin{equation} 
\vspace{-1mm}
E = mc^{2}
\end{equation}
Each collision is referred to as an event. In every event, the position (\emph{hits}) and momentum of newly created particles are measured by the detectors which surround the interaction region. The raw data from the detector is used to reconstruct the collision events in a process known as event reconstruction, which involves identifying the particles, their energy, momentum, trajectory, and their point of origin, hence measuring the process that took place at the collision \cite{7}. 
The first run of the LHC (2009) achieved 7 TeV of collision energy, subsequently increased to 13 TeV during Run 2 (2015). The LHC was shutdown in 2018 for additional upgrades and is scheduled to resume operations in 2022 (Run 3) with an expected collision energy of 14 TeV \cite{8}.

The amount of data generated at each event is expected to increase as HEP\footnote{High Energy Physics} experiments are upgraded, since more matter is created at higher energies. The four-major experiments at the LHC during Run 3 are expected to generate data at a rate of $\mathcal{O}(10)$ Terabit/s \cite{9}. Storing all of this data in real-time for post-analysis is logistically infeasible, which is why experiments at the LHC use a trigger system. By partially reconstructing the events in real time, a trigger system selects and stores interesting datasets for detailed post analysis. The LHC is expected to generate 40 million bunch proton-proton (pp) collisions per second (40 MHz). Hardware level triggers in the ATLAS and CMS aim to reduce this throughput rate down to $\mathcal{O}(1)$ KHz \cite{7}. In contrast, the upgraded LHCb experiment will use a software based trigger running entirely on GPUs operating at the full bunch collision rate \cite{10}.

\begin{figure}[H]
\includegraphics*[width=1\linewidth,clip]{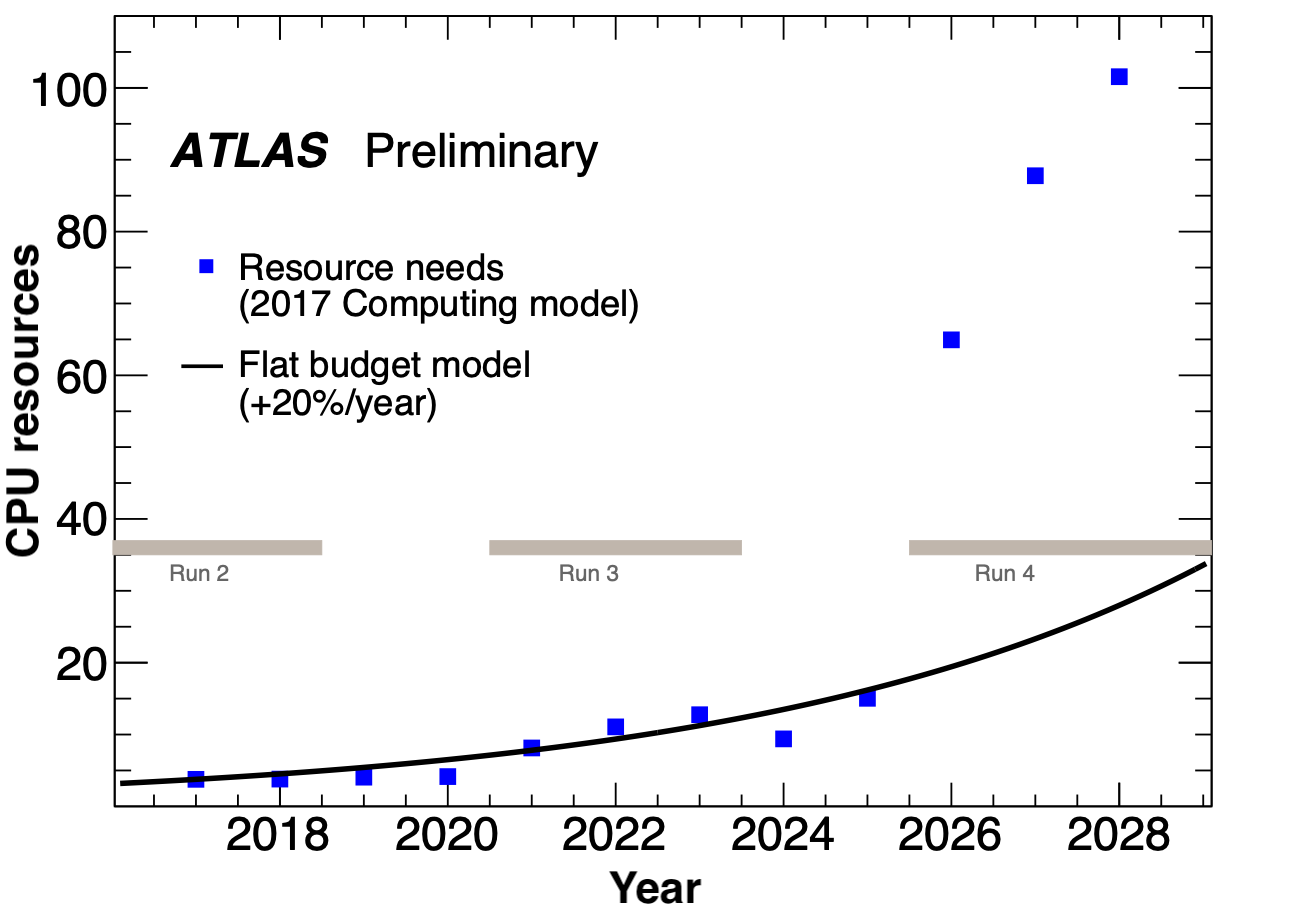}
\caption{Estimated CPU resources needed (blue points) by the ATLAS experiment from the years 2018 to 2028. The solid line represents the amount of resources expected to be available assuming a flat funding scenario of 20\% per year. Other experiments at the LHC follow similar resource-forecast trend. Figure taken from \cite{11}} \label{fig:3}
\end{figure}

\setlength\belowcaptionskip{-1.1ex}
\begin{figure}[H]
\includegraphics*[width=0.9\linewidth,clip]{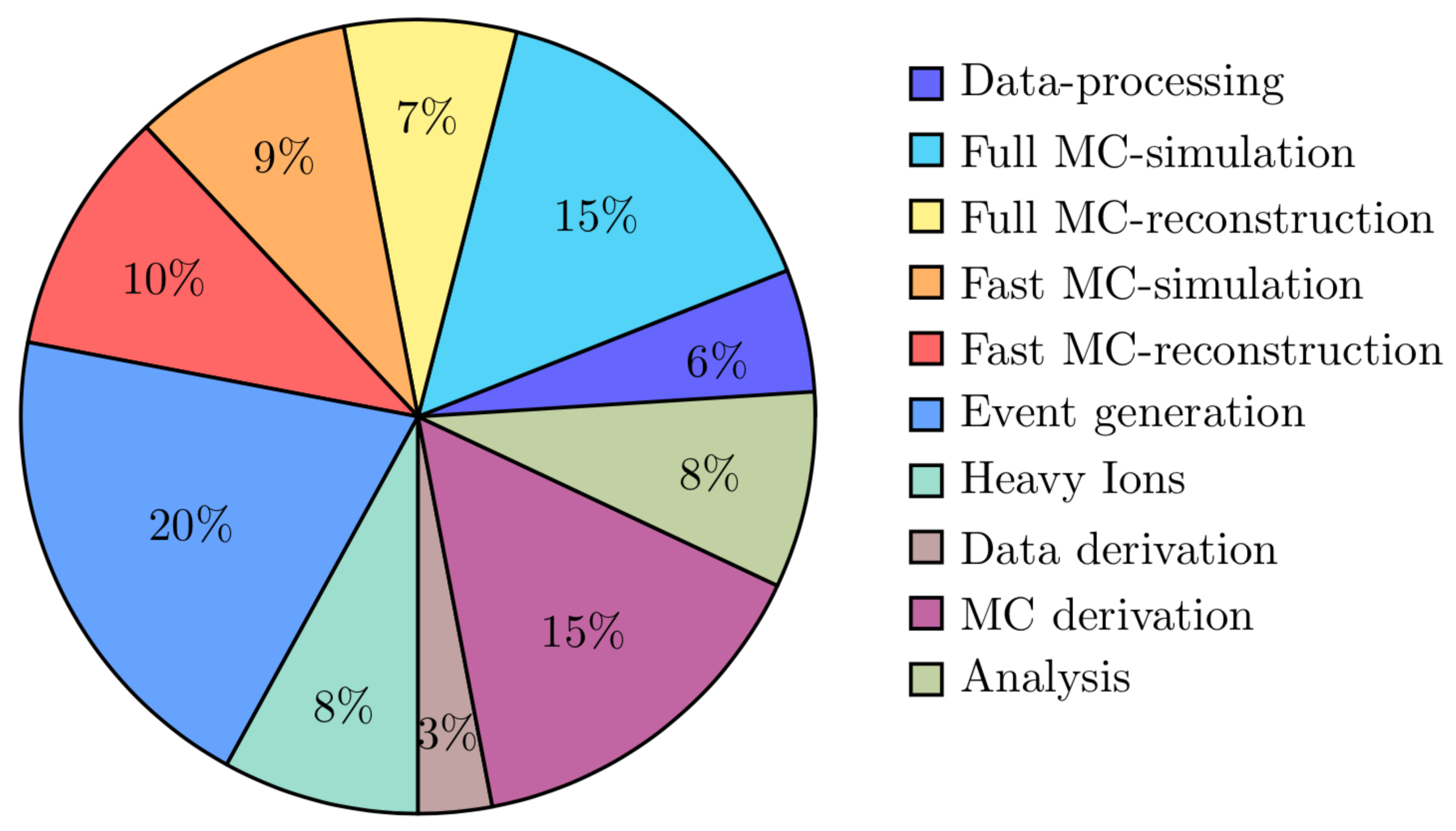}
\caption{A pie chart of projected CPU resources (by computational task) required by ATLAS in 2030.} \label{fig:2}
\end{figure} 
In addition to the immense computing resources required by the trigger, HEP experiments also ubiquitously use Monte Carlo (MC) simulations, mainly for modelling the collision events, detector response, detector calibration, and more (see figure \ref{fig:2} for CPU resource allocation by task). A steep rise in computing resources is needed to keep up with the exponentially increasing size of particle physics datasets. Future experiments would not only face an engineering challenge, but a computational one as well since these experiments would be limited by how efficiently the computing resources can be used. A four-fold shortage in computing power (figure \ref{fig:3}) is forecasted by 2030 \cite{11}. This is because, although the transistor density of a CPU\footnote{Central Processing Unit} has been increasing over time; obeying Moore's law (albeit is decelerating), the single-core CPU performance has plateaued since the mid-2000s due to the constraints on power density \cite{12}.
 \vspace{-2mm}
\setlength\belowcaptionskip{-2ex}
Due to these performance constraints, particle physics experiments such as ATLAS, COMET, ALICE, CMS, and LHCb \cite{13,14,15} are investigating and implementing alternate computing techniques and multi-core architectures (\textbf{i.e} software based GPU trigger at LHCb) to cope up with the ever increasing computing demands of HEP experiments. These include heterogeneous systems where computing architectures such as the GPUs\footnote{Graphics Processing Unit} and FPGAs\footnote{Field programmable gate arrays} work in conjunction with the CPUs. A modern CPU contains $\mathcal{O}(10)$ powerful cores built on the MIMD\footnote{Multiple Instruction, multiple Data} design, whereas the GPU contains $\mathcal{O}(1000)$ simpler cores built on the SIMD\footnote{Single instruction, multiple data} design, making the GPU ideal for running multiple identical computations in parallel. On the other hand, FPGAs are integrated circuits that are not hard etched; unlike CPUs and GPUs, this allows FPGAs to be reprogrammed for speciﬁc computations, often resulting in far superior performance compared to its traditional counterparts.

This urgency of coping up with an ever-increasing need for computing resources in HEP experiments plays a central role in this research project, which aims to:
\vspace{-2mm}
\begin{itemize}[leftmargin=+0.4cm]
\itemsep-0.1em 
\item[--] Give an overview of computing techniques used in HEP experiments (with a focus on the LHCb experiment).
\item[--] Develop HEP algorithms, using traditional and machine learning based techniques.
\item[--] Implement and benchmark these algorithms on multiple computing architectures such as: CPUs, GPUs, and a new kind of processor: the IPU\footnote{Intelligence Processing Unit}. 
\end{itemize}
The report is organised as follows: section 2 aims to give an overview of the physics at the LHCb. Section 3 describes the computing architectures used in this research project, followed by the description of the trigger system in section 4. Section 5 outlines different stages of event reconstruction, followed by section 6, which describes the algorithms developed in this study and their performance. Finally, section 7 discusses the implications of these results, the limitations, further work that can be carried out, and section 8 concludes this study.

\section{Physics at the LHCb}
The LHCb experiment is designed to investigate decay channels and oscillations of beauty and charm hadron systems with a particular focus on CP violating phenomena, as well as searches for anomalies in rare decays which indicate physics beyond the standard model. 
\vspace{-4mm}
\subsection{CP violation}
The standard model of particle physics (SM) encapsulates our current best understanding of three out of the four fundamental forces in the universe (weak, strong, electromagnetic, excluding gravity). The SM classifies elementary particles according to their charges and describes their fundamental interactions. In particle physics, the charge conjugation parity symmetry states that the laws of physics must remain invariant when a system's spatial coordinates are flipped (parity inversion) eq(2) and the particle is replaced by its antiparticle (charge conjugation) eq(3). This symmetry was first observed to be broken in 1964 in the decays of neutral kaons \cite{16}. 
\begin{align}
\hat{P}\left (  \hat{x},\hat{y},\hat{z} \right ) \rightarrow \left (  -\hat{x},-\hat{y},-\hat{z} \right )
\end{align}
\vspace{-6mm}
\begin{align}
\hat{C} \ket{\Psi}_{e^{-}} = \ket{\Psi}_{e^{+}}
\end{align}
The three generations of quarks in the SM naturally generate CP violating phenomena in both weak and strong interactions. 
CP violation in weak interactions is described via a complex unitary matrix known as the Cabibbo Kobayashi Maskawa matrix (CKM matrix) \cite{17}. Furthermore, CP violation is also extremely relevant in cosmology since it can describe the matter-antimatter asymmetry observed in our universe. However, the level of CP violation needed to generate this asymmetry cannot be accounted for by the standard model, which indirectly points towards BSM \cite{18}.

\vspace{-4mm}
\subsection{The Cabibbo-Kobayashi-Maskawa matrix}
The CKM matrix determines mixing between the three families of quarks (by mapping orthogonal basis to orthogonal basis, hence unitary). The matrix element (V\textsubscript{i,j}) describes the relative strength of transition (i$\mapsto$j) between quarks.
\begin{equation}
V\textsubscript{CKM} = 
\begin{pmatrix}
V\textsubscript{ud} & V\textsubscript{us} & V\textsubscript{ub} \\
V\textsubscript{cd} & V\textsubscript{cs} & V\textsubscript{cb} \\
V\textsubscript{td} & V\textsubscript{ts} & V\textsubscript{tb}
\end{pmatrix}
\end{equation}

The CKM matrix is defined by four parameters, and the Wolfenstein parameterization is conventionally used due to its convenience. The third-order expansion of the CKM matrix in terms of the Wolfenstein parameters is as follows:
\begin{equation}
V^{(3)}\textsubscript{CKM} = 
\begin{pmatrix}
1-\lambda^{2}/2 & \lambda & A\lambda^{3}(\rho-i\eta) \\
-\lambda & 1-\lambda^{2}/2 & A\lambda^{2} \\
A\lambda^{3}(1 - \rho-i\eta) & -A\lambda^{2} & 1
\end{pmatrix}
\end{equation}

\begin{figure}[H]
\includegraphics*[width=1\linewidth,clip]{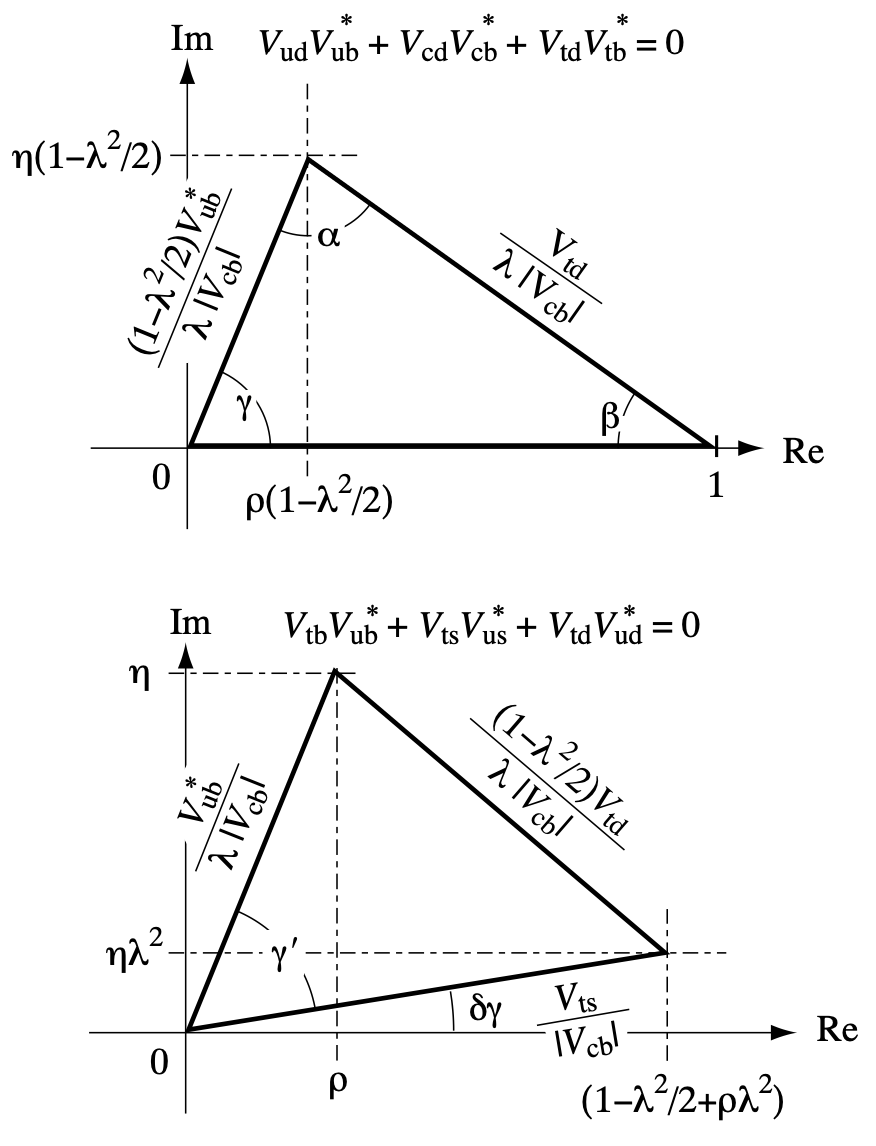}
\caption{Two complex triangles representing the two unitary conditions of the CKM matrix relevant to the B-sector. Figure taken from Ref \cite{20}.} \label{fig:4}
\end{figure}

As mentioned earlier, CP violation was first observed in the Kaon sector; however,
this sector is quite limited in terms of CP violating channels. In contrast, the SM predicts many CP violating modes in the B-meson sector at energy scales accessible by the LHC (larger phase space), making the B-sector very appealing to study CP violating phenomena \cite{19}. 
The unitarity of the CKM matrix generates six unitary conditions, which can be drawn as triangles \cite{20}. Figure \ref{fig:4} depicts two such triangles representing the two conditions relevant to the B-sector.
The angles of these triangles can be measured either indirectly (by measuring the length of the sides) or directly (from CP asymmetry in B-meson decays). A disagreement between these two methods would implicate BSM \cite{20}. The matrix elements V\textsubscript{cb} and V\textsubscript{ub} are deduced by observing various B-meson decay modes, and the elements  V\textsubscript{td} and V\textsubscript{ts} are determined by measuring the frequency of $B$$^0_d$ -$\overline{B}$$^0_d$ and $B$$^0_s$ -$\overline{B}$$^0_s$ oscillations respectively. These matrix elements are used to derive the Wolfenstein parameters, which ultimately defining the angles $\alpha$, $\beta$, $\gamma$, and $\delta$\textsubscript{$\gamma$} \cite{20}. The following list contains a few examples of B-meson decays which allow these angles to be directly measured \cite{21}:
\begin{enumerate}
  \item $\beta + \gamma$ from $B^{0}_{d} \mapsto \pi^{+} \pi^{-}$
  \item $\beta$ from $B^{0}_{d} \mapsto J/\psi K_{S}$
  \item $\gamma - 2\delta_{\gamma}$ from $B^{0}_{s} \mapsto D^{\pm}_{s} K^{\mp}$
  \item $\delta_{\gamma}$ from $B^{0}_{s} \mapsto J\psi \phi$
  \item $\gamma$ from $B^{0}_{d} \mapsto \bar{D}^{0}K^{*0},D^{0}K^{*0},D_{1}K^{*0}$
\end{enumerate}

\setlength\belowcaptionskip{-1.4ex}

\subsection{BSM signatures}
Comparing the SM predictions with precise measurements would  rule out/account for new flavour physics models \cite{22}. For example: new flavour suppressing models could have a significant impact on $B$$^0_s$ -$\overline{B}$$^0_s$  and $B$$^0_d$ -$\overline{B}$$^0_d$ oscillations. Furthermore, new physics could potentially have a large effect on b $\mapsto$ s transitions which would affect: B\textsubscript{s} $\mapsto$ J/$\psi$($\mu$$\mu$)$\phi$, or the very rare decay: B\textsubscript{s} $\mapsto$ $\mu$\textsuperscript{+}$\mu$\textsuperscript{-} via the Higgs penguin diagram \cite{23}. A recent analysis of the data from the LHCb experiment deviates from the lepton flavour universality predicted by the SM. The measured quantity of interest is expressed as $R_{K}$:

\begin{equation}
R_{K} = \frac{\ss(B \mapsto K^{*}\mu^{+}\mu^{-})}{\ss(B \mapsto K^{*}e^{+}e^{-})}  \stackrel{SM }{\cong} 1
\end{equation}

The analysis estimates $R_{K}$ to be $0.846^{+0.042}_{-0.039}$, which is a 3.1 sigma deviation from the SM \cite{24}, suggesting that the forces of nature treat muons and electrons differently, hinting towards BSM. Additional data (Run 3) is required to classify this as a genuine deviation (5 sigma standard). Nonetheless, these results have sparked tremendous excitement in the particle physics community.

\begin{figure*}
\captionsetup{singlelinecheck = false, justification=justified}
  \includegraphics[width=0.7\linewidth,clip]{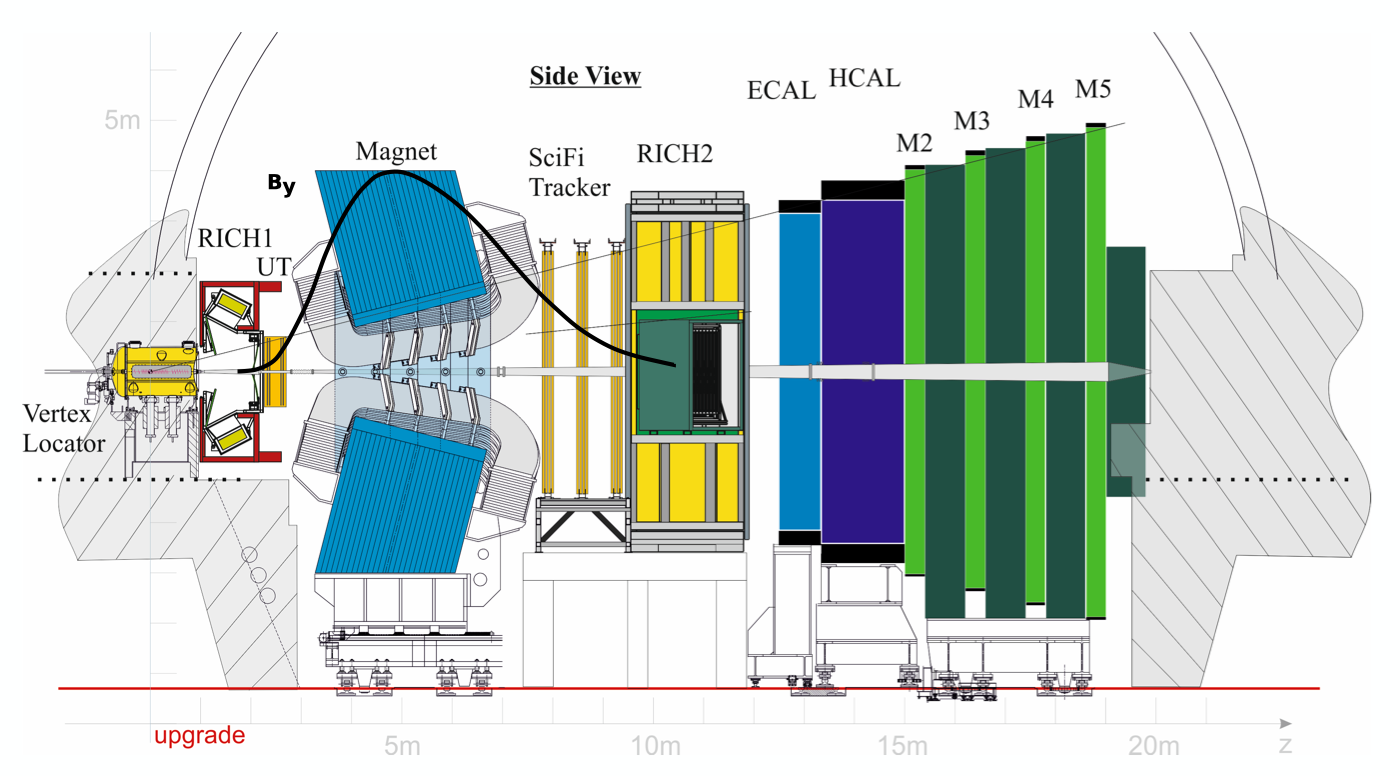}
  \caption{Schematic of the upgraded LHCb detector. Figure taken from Ref \cite{10}.}\label{fig:5}
\end{figure*}

\newpage

\subsection{Schematic of the LHCb detector}
The LHCb is a single-arm forward spectrometer that measures proton-proton interactions using a high-precision tracking system consisting of a silicon-strip vertex detector (Velo) surrounding the interaction region, and the upstream tracker (UT) placed before the magnet. Placed after the magnet are the scintillating fibres (SciFi), the Ring Imaging Cherenkov (RICH, used for hadrons Identification), and the Muon chambers. Charged particles in the Velo detector transverse in straight lines, which get bent by the magnet in the UT and SciFi detectors (bending determines the momentum). This tracking system measures the momentum ($p$) of charged particles with an uncertainty varying between 0.5\% at low momentum to 1.0\% at 200 GeV/c. The impact parameter (minimum distance of a track to a proton-proton collision) is measured with a resolution of (15 + 29/$p_{T}$) $\mu$m, where $p_{T}$ is the component of the momentum relative to the beamline.

Making precise experimental measurements of B-sector interactions requires differentiating kaons from pions since the decay modes of interest are heavily contaminated by: B\textsubscript{d} $\mapsto$ K\textsuperscript{$\pm$}$\pi$\textsuperscript{$\mp$}, B\textsubscript{s} $\mapsto$ K\textsuperscript{$\mp$}$\pi$\textsuperscript{$\pm$}, and B\textsubscript{s} $\mapsto$ K\textsuperscript{$\mp$}K\textsuperscript{$\pm$} decays \cite{20}. Furthermore, measuring the rapidly oscillating B-meson systems requires excellent proper-time resolution. The LHCb detector's excellent momentum and vertex resolution satisfies both of these conditions \cite{25}. At its core, the LHCb experiment takes advantage of the sheer number of B-hadrons produced at the LHC, allowing it to make precise measurements of rare decays and meson oscillations in the B-sector. 

\section{Computing architectures}

As discussed in section 1, the size of HEP datasets is expected to increase exponentially over time. Additionally, the plateauing single-core CPU performance means that CPU exclusive systems cannot cope with the ever-growing computing demands of HEP experiments \cite{11}, which is why many such experiments are investigating alternate computing architectures \cite{13,14,15}. This section aims to give an overview of the four computing architectures used in this research project. 
The performance of these architectures (excluding the TPU) is investigated for both ML and non-ML based algorithms. The use of a TPU is limited to train/validate neural networks. All algorithms created in this research project are written in Python using the TensorFlow framework, which is supported by the architectures listed in table 1.

\begin{table}[H]
\begin{adjustbox}{width=\columnwidth,center}
\begin{tabular}{cccccc}
\textit{Type} & \textit{Name} & \textit{Cores} & \textit{Memory} & \textit{Clock speed} & \textit{Power consumption} \\ \hline
\multicolumn{1}{|c|}{\textit{CPU}} & \multicolumn{1}{c|}{\textit{Intel Xeon Silver 4215R}} & \multicolumn{1}{c|}{\textit{8}} & \multicolumn{1}{c|}{\textit{396000 MiB}} & \multicolumn{1}{c|}{\textit{3.2-4.0 GHz}} & \multicolumn{1}{c|}{\textit{130 W}} \\ \hline
\multicolumn{1}{|c|}{\textit{GPU}} & \multicolumn{1}{c|}{\textit{NVIDIA T4}} & \multicolumn{1}{c|}{\textit{2560}} & \multicolumn{1}{c|}{\textit{16000 MiB}} & \multicolumn{1}{c|}{\textit{8.1 TFLOPS}} & \multicolumn{1}{c|}{\textit{70 W}} \\ \hline
\multicolumn{1}{|c|}{\textit{IPU}} & \multicolumn{1}{c|}{\textit{Graphcore Colossus GC2}} & \multicolumn{1}{c|}{\textit{1216}} & \multicolumn{1}{c|}{\textit{286 MiB}} & \multicolumn{1}{c|}{\textit{31.1 TFLOPS}} & \multicolumn{1}{c|}{\textit{120 W}} \\ \hline
\multicolumn{1}{|c|}{\textit{TPU}} & \multicolumn{1}{c|}{\textit{TPU v2-8}} & \multicolumn{1}{c|}{\textit{8}} & \multicolumn{1}{c|}{\textit{64000 MiB}} & \multicolumn{1}{c|}{\textit{180 TFLOPS}} & \multicolumn{1}{c|}{\textit{40 W}} \\ \hline

\end{tabular}
\end{adjustbox}
\caption {Key specifications of the processors used in this study}
\end{table}

\subsection{Central Processing unit}
A modern CPU contains about $\mathcal{O}(10)$ high-frequency (complex) cores primarily designed for sequential serial processing. Each core can work independently on various logical calculations (multiple instructions multiple data) at the same time \cite{26}. The MIMD paradigm allows the CPU cores to deal exceptionally well with control-dominated techniques such as branch prediction, branch speculation, and out-of-order execution with extremely low latency. Furthermore, the MIDM design allows the CPU to excel in general-purpose computing. For example, a general algorithm (consisting of multiple different tasks) can be broken down and executed on individual cores. CPU cores share a layer of high-speed memory known as cache, which enables them to share the results of their individual computations if needed. The smaller yet complex core count means that CPUs deal well with dynamic workflows at the expense of aggregate arithmetic throughput per area of silicon \cite{27}.

\subsection{Graphics Processing Unit}
Unlike the CPU, the GPU contains about $\mathcal{O}(1000)$ simpler cores optimized for parallel processing. A kernel is an algorithm that is executed on the GPU. Each kernel is launched with a certain number of threads that execute the same instruction on multiple parts of the data simultaneously (SIMD design). Threads are grouped together into blocks, forming a grid configuration. Threads within the same block share memory (allowing them to share data and synchronize), whereas threads belonging to different blocks cannot communicate \cite{10}; this is known as the SIMT\footnote{Single Instruction, Multiple Threads} hierarchy. 
This arrangement of clustering $\mathcal{O}(1000)$ simpler cores allows the GPU to excel at dense, regular, numerical, data-dominated workflows \cite{27} such as Machine learning, computer-generated imagery (CGI), video editing, and more. Given a regular parallelizable task, GPUs tend to offer much larger arithmetic throughput per area of silicon compared to the CPU because they dedicate a higher fraction of their silicon to arithmetic units.

On the other hand, the simpler design of the GPU cores means that they cannot offer techniques such as branch speculation and out-of-order execution to deal with irregular computations \cite{27}. The SIMT architecture of a GPU is prone to hefty performance penalties due to a phenomenon known as warp divergence \cite{28}, which occurs when multiple threads running the same instructions diverge in their control flow (occurs in the presence of conditional statements). 

\vspace{-3mm}
 \subsection{Intelligence Processing Unit}
Built by Graphcore (a Bristol based Semiconductor company), the IPU contains about $\mathcal{O}(1000)$ cores called \emph{tiles}, which are complex enough to compute separate instructions independently (MIMD design) \cite{27}. The IPU completely disregards the shared memory architecture of the CPU (cache) or the GPU (block memory). The novel memory architecture of the IPU assigns each tile with 256 KiB of memory implemented as S-RAM\footnote{Static random access memory} (eliminating memory access bottlenecks associated to shared memory architectures). The IPU-Exchange\textsuperscript{TM}  allows for point to point communication between the tiles. 
The resulting memory performance is comparable to that of CPU caches and is superior to the GPU memory in terms of latency and bandwidth (45 TB/s bandwidth and 6x latency boost over DRAM\footnote{Dynamic random access memory}) \cite{29}. This memory architecture allows a system with multiple IPUs to effectively act as a single virtual IPU device with no extra development effort, which is not the case when clustering multiple CPUs or GPUs. This scalability sets the IPU apart from the competition.
Furthermore, the IPU is immune to effects like warp divergence due to its MIMD architecture. For this reason, the IPU is more efficient than the GPU in computations involving heavy use of control-flows and workflows which require random, sparse, irregular data access. 
\vspace{-3mm}
\subsection{Tensor Processing Unit}
The TPU is a new type of coprocessor designed from the ground up by Google and is custom-tailored towards machine learning applications written in TensorFlow. The TPU is a multi-core ASIC\footnote{application-specific integrated circuit} where each core is assigned 8 GiB of high-bandwidth memory. Each TPU core contains scalar, vector, and matrix units (MXU), where the MXU holds the majority of the computing power (since matrix operations are ubiquitous in ML applications) \cite{31}. The MXU is built for high volume of low precision computation (\textbf{e.g} 16-bit operations), allowing it to train large ML models much faster \cite{32}. The TPU is accessible via Colab: a web-based IDE created by Google for machine learning research.
\vspace{-3mm}
\section{Trigger system}
The Run 3 of the LHCb would operate at an instantaneous luminosity of 2 $\times$ 10\textsuperscript{33} cm\textsuperscript{-2}s\textsuperscript{-1}, a factor of five increase over the Run 2. The LHCb's particle-identification system, tracker detectors, readout, and data-acquisition systems are being replaced to accommodate this increase. This will enable the experiment to readout the increased number of collisions directly into the software-based trigger system (bypassing the deadtime caused by the hardware level trigger) \cite{33}.

The 30MHz collision rate translates to a data rate of 40 Tbit/s generated by the LHCb detectors. At this data rate, it is more feasible to categorise the signal according to the physics requirements instead of rejecting background from the signal \cite{34}. This approach requires considerably more information than the hardware level trigger used in Run 2 can provide. LHCb and ALICE are the first experiments at the LHC to exclusively use a software-based High-level trigger (HLT).

\begin{figure}[H]
\includegraphics*[width=1\linewidth,clip]{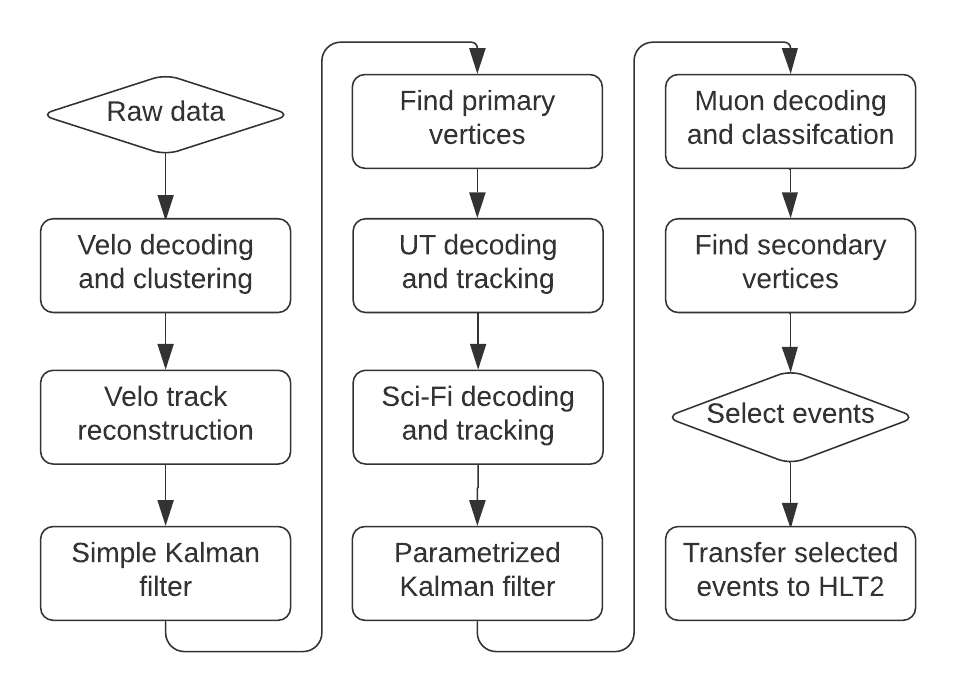}
\caption{Flowchart representing the HLT1 sequence written in CUDA (Nvidia's API) is executed on 500 GPUs. The event building servers copy the raw assembled data on to the GPUs, which perform the following HLT1 sequence and transfer the output back to the host CPUs. This output is packaged and sent to the event filter farm for the HLT2 sequence. The rhombi represents the start and endpoints of the sequence, and the rectangles represent the algorithms processing data. Flowchart Ref \cite{10}.} \label{fig:6}
\end{figure}

The trigger consists of two stages: The first stage, called HLT1, performs basic track reconstruction using a subset of algorithms used in Run 2. The detector alignment and calibration constants from the previous `runs' are used \cite{10}. The most valuable tracks are reconstructed first before performing basic candidate selection (figure \ref{fig:6} represents the full HLT1 sequence) \cite{25}. The following signatures are used to select candidate HL1 tracks \cite{10}:
\begin{itemize}[leftmargin=+0.4cm]
\itemsep-0.4em 
\item[--] Single track:  $p_{T} > 1$ GeV.
\item[--] Double track: $p_{T} > 0.7$ GeV each.
\item[--] High $p_{T}$ muon: Muon track with $p_{T} > 10$ GeV.
\item[--] Displaced dimuon: A displaced dimuon vertex with $p_{T} > 0.5$ GeV for both tracks or  High-mass dimuon with $p_{T} > 0.75$  GeV for both tracks
\end{itemize}

The second stage, called HLT2, performs offline quality event reconstruction by aligning and calibrating the detector in near real-time, \textbf{i.e}:
track reconstruction, particle identiﬁcation, and track fitting on the events selected by the HLT1 \cite{35}. Besides selecting events, the HLT2 stage identifies decays of interest and assigns them to one of the primary tracks. The HLT2 outputs the reconstructed events an order of magnitude smaller than the raw input data, reducing the data rate to 80 Gbit/s \cite{36}.

\begin{figure}[H]
\includegraphics*[width=1\linewidth,clip]{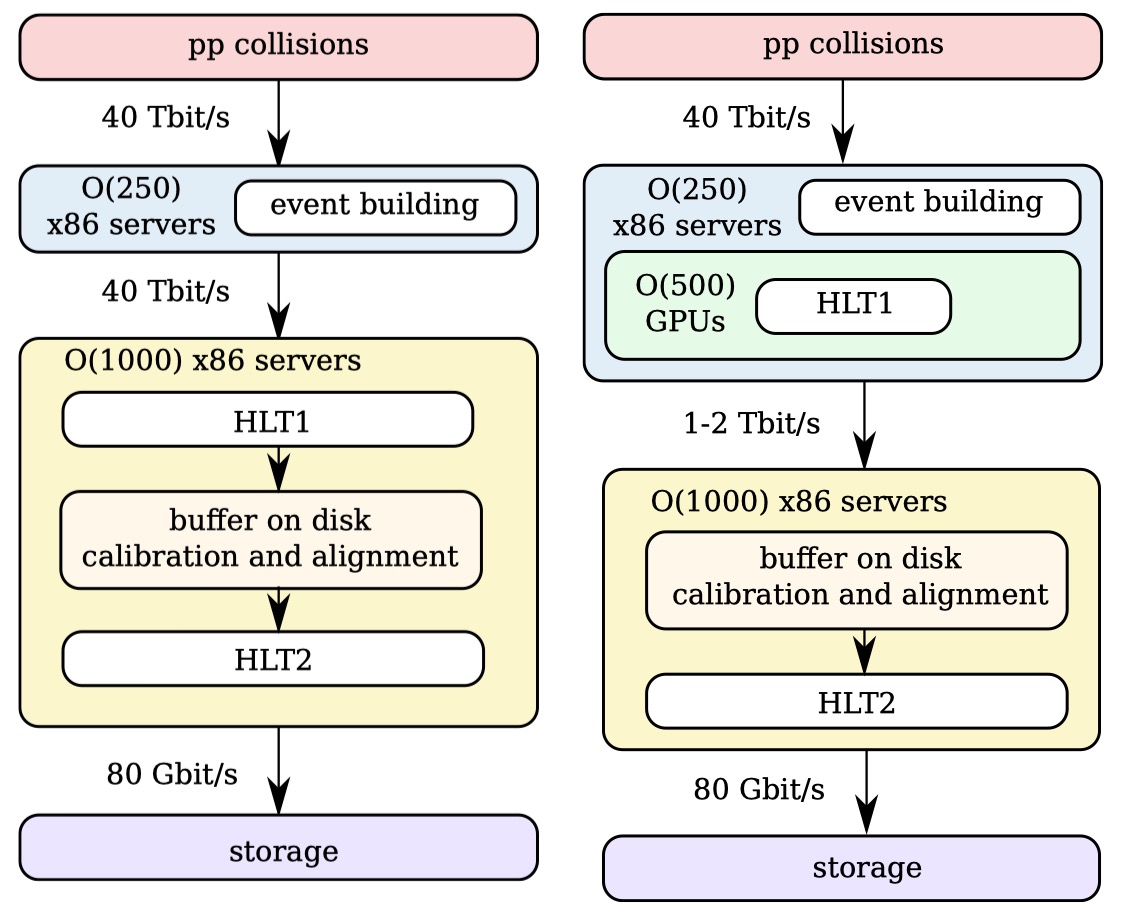}
\caption{Schematic of the data acquisition system at the LHCb. The image on the left represents the data flow during Run 2 consisting of $\mathcal{O}(1000)$ x86 servers dedicated for HLT 1 \& 2 stages. The image on the right represents the data flow during Run 3, where the GPU-HLT1 stage is executed within the event building servers \cite{15}.} \label{fig:7}
\end{figure}

The majority of the experimental data at the LHCb is generated by recording \enquote{hits} left by the particles on the tracking detectors. Data from different sub-detectors is collected by 500 FPGA cards and combined into complete events on 250 event building x86 servers  \cite{10}. During run 2, the assembled events were transferred to the event filter farm (containing 1000 x86 servers), where the HLT1 and HLT2 stages were executed. The Allen proposal (Run 3) exploits the fact that track reconstruction is an inherently parallel problem and makes use of the multicore SIMD architecture of the GPU to reconstruct events \cite{10}. 
This implementation requires about 500 consumer-grade GPUs. This fits right into the data-acquisition system of the LHCb (since each FPGA card can host two GPU cards via two PCI slots), allowing the LHCb to execute the HLT1 stage within the event building servers (reducing the data rate and the networking cost by a factor 30-60). See Fig \ref{fig:7} for a comparison between Run 2 and Run 3.

Presently, only the ALICE experiment uses GPUs in its trigger to reconstruct tracks (although not used for event selection or data reduction) from individual sub-detectors (albeit at a lower data rate of 5 Tbit/s) \cite{37}. Other GPU based proposals intend to: analysis data from a single sub-detector, perform parallel track reconstruction and event selection by looking for specific physics signatures in the dataset \cite{38,39,40}.

\section{Event reconstruction}
Event reconstruction plays a crucial role in HEP experiments. This stage involves transforming the raw data (particle hit positions, hit energies, ionization, times stamps) recorded by the detector into complete events, used to perform the physics analysis. This analysis crucially relies on \cite{7}:
\begin{itemize}[leftmargin=+0.4cm]
\itemsep-0.4em 
\item[--] Identifying which particles were created.
\item[--] knowing where the particles were created.
\item[--] Identifying the parent particles.
\end{itemize}
This information is obtained by performing the following \cite{7}: 
\begin{itemize}[leftmargin=+0.4cm]
\itemsep-0.4em 
\item[--] Track-reconstruction: rebuilding particle trajectories
\item[--] Vertexing: Group particles into vertices.
\item[--] Particle ID: Identifying the type of particles
\end{itemize}

\subsection{Track reconstruction}
Reconstructing charged particle tracks consists of two stages: pattern recognition (finding which detector hits belong to the same track) and track fitting (parametrising the track). This subsection gives an overview of the algorithms used for reconstructing tracks in different sub-detectors at the LHCb. 
\begin{figure*}
\captionsetup{singlelinecheck = false}
  \includegraphics[width=1\linewidth,clip]{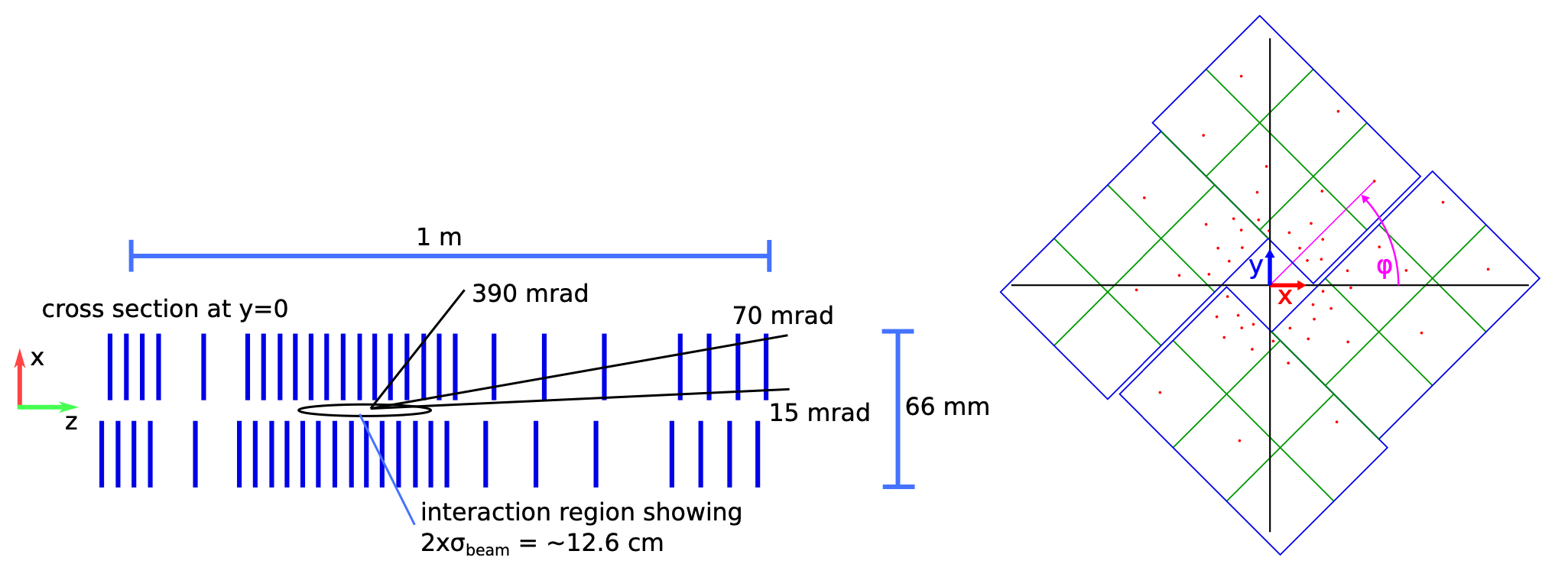}
  \caption{A representation of the Velo detector. Figure on the left represents the layout of the Velo detector (\textbf{i.e}: 26 silicon silicon-strip modules near the region of interaction from where the tracks (lines) emerge), the (x,y,z) coordinate system is aligned with the beamline with the z axis pointing downstream. The figure on the right represents a single silicon module, where each square is a 14mm x 14mm sensor with a resolution of 256 x 256 pixels. Charge particles transverse the Velo detector at constant angles  ($\varphi$). Figure from Ref \cite{44}.} \label{fig:8}
\end{figure*}

\subsubsection{Vertex locator (Velo)}
The Velo detector consists of 52 silicon-strip sensors surrounding the interaction region (26 sensors on each side), and it establishes the first stage of track reconstruction \cite{41} (see figure \ref{fig:8}). Charged particles created in the interaction region transverse through the Velo detector, leaving hits in a straight line (due to the absence of a magnetic field) towards other detectors (moving left to right). The main objective here is to reconstruct the initial track segments and their primary vertices (\textbf{i.e} points in space where (pp) interactions have occurred) \cite{42}. The reconstructed track segments are then extended to other subsequent LHCb detectors \cite{10}.

\begin{figure}[H]
\includegraphics*[width=1\linewidth,clip]{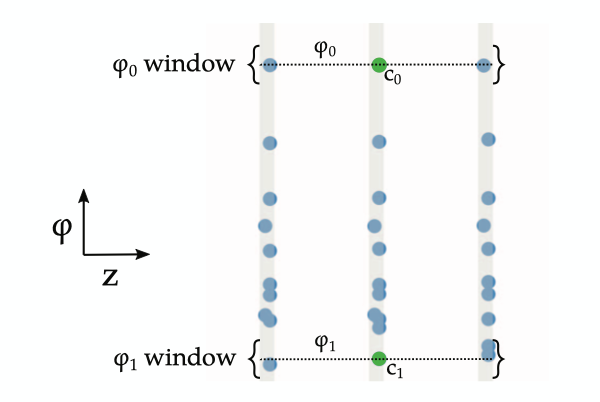}
\caption{Depiction of three adjacent Velo modules. The hits in each module are transformed into their respective $\varphi$ values. Hits in the middle module $c\textsubscript{0}$ and $c\textsubscript{1}$ correspond to the angles $\varphi\textsubscript{0}$ and $\varphi\textsubscript{1}$. The $c\textsubscript{0}$ hit has a compatible hit in the preceding and following module, forming a track seed (triplet). Whereas $c\textsubscript{1}$ has a compatible hit in the preceding module and two compatible hits in the following module (resulting in two triplets). In such a case, the better performing track seed is selected (via $\chi^{2}$ metric).  Figure Ref \cite{43}.} \label{fig:9}
\end{figure}

Reconstruction at the Velo employs the \emph{search by triplet} algorithm, which exploits the fact that charged particles in the Velo detector travel in straight lines (in constant angles $\varphi$ within a cylindrical coordinate system) \cite{43}. This implies that hits in adjacent silicon modules belonging to `similar' tracks would have similar $\varphi$ values (arctangent of a hit coordinate relative to the origin). The algorithm computes the arctangent of every hit in every module and sorts the hits in ascending values of $\varphi$. This algorithm operates on three detector modules at a time (in parallel on GPUs), allowing it to create line segments by forming seeds of three hits. It does this by comparing hits in the adjacent (\textbf{i.e} previous and next) modules whose $\varphi$ values fall within a window of acceptance \cite{43}. Since identifying track seeds is an inherently parallel problem, multiple GPU threads are assigned to compute different track seeds (differing in $\varphi$ values) in the same set of modules. Some of the hits may form multiple track seeds (see figure \ref{fig:9}), resulting in a workload imbalance; for this reason, multiple threads can be assigned to the same hit to balance the workload. Once the threads have created their respective seeds, the best seed ($\chi^{2}$ defines the seed quality) passing through each `middle' hit is selected. 

The resulting track seeds are forwarded to the next module, and a binary search is performed to find hits closest to the forwarded track segments. Hits below a certain distance threshold are appended to the existing track segment. This process is repeated over all Velo silicon modules. The worst-case complexity of this algorithm is $\mathcal{O}(m^{2} \cdot n^{ } \cdot log(n))$ where m and n are the number of modules and the average number of hits in each module respectively \cite{43}.

\setlength\belowcaptionskip{-1.5ex}

For context, it is worth mentioning the existence of other tracking techniques:
\begin{itemize}[leftmargin=+0.4cm]
\itemsep-0.4em 
\item[--] Full combinatorial: This generates every possible combination of hits in different modules to generate tracks. This approach is computationally expensive since the complexity scales as $\mathcal{O}(n!)$ \cite{7}, where n is the average number of hits in each module.
\item[--] Hough transform: This involves recasting the hits into lines in the Hough space. Hits belonging to a common track in real space would form a cluster of intersecting lines in the Hough space \cite{45}. A histogram is used to identify such clusters, allowing to group compatible hits together. The complexity scales as $\mathcal{O}(n \cdot m^{2})$ where n and m are the the total number of hits and the number of bins in the histogram respectively.
\item[--] Artificial retina (discussed later): This involves building a discrete grid (heatmap) in the track parameter space. High-intensity cells in this grid describe the track parameters of a dataset, and hence the compatible hits \cite{46}. The complexity scales as $\mathcal{O}(n \cdot m^{2})$ where n and m are the total number of hits and the number of discrete cells in the parameter space respectively.
\end{itemize}
\vspace{-4mm}

\subsubsection{Upstream tracker (UT)}
\begin{figure}[H]
\includegraphics*[width=0.85\linewidth,clip]{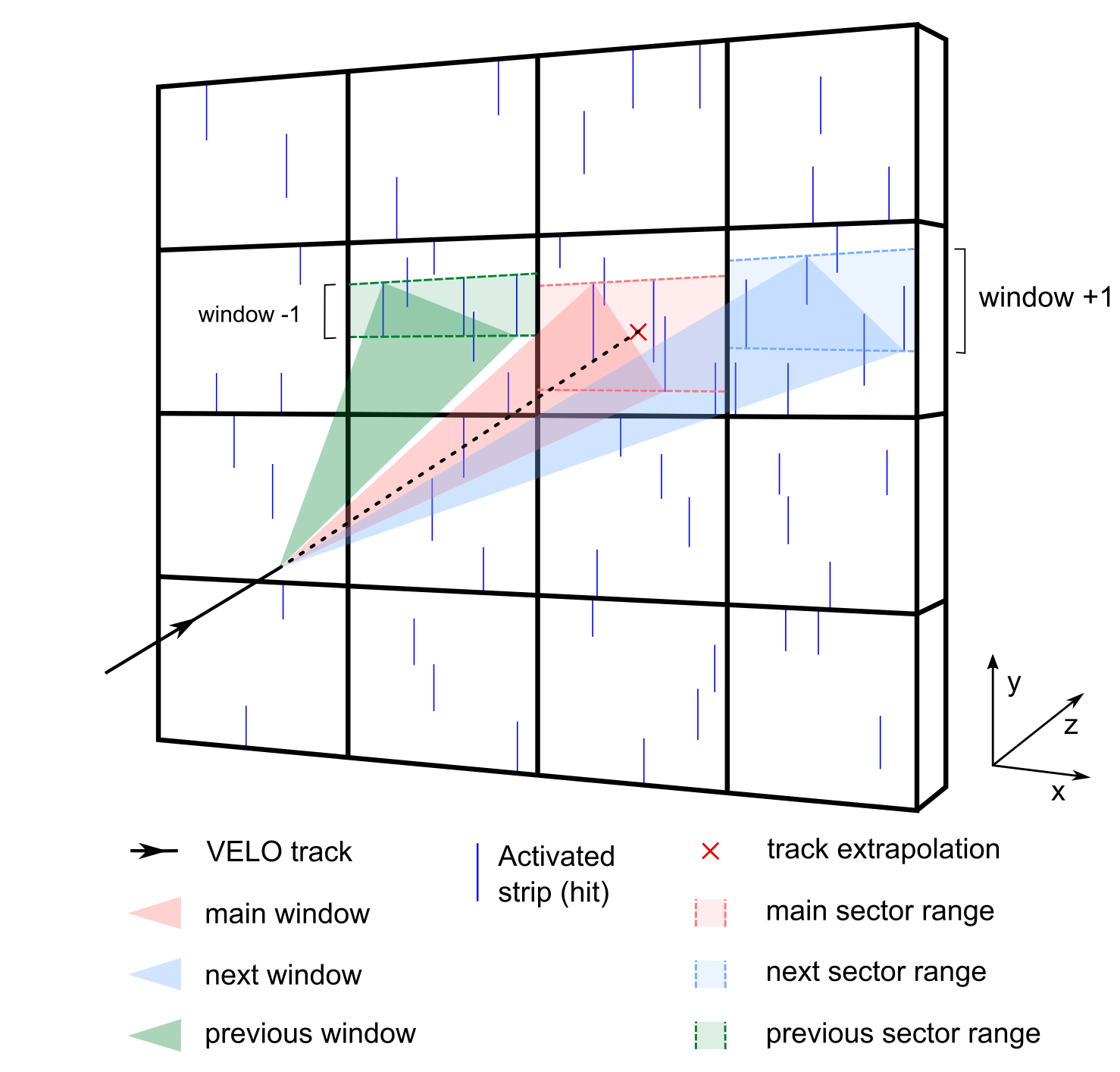}
\caption{Depiction of a single UT strip. The UT hits are segregated into search windows, and the Velo tracks are extended to the UT. The hits within the search window are compared with the against the Velo track tolerance. Each search window processed in parallel, and the best compatible hits are appended to the Velo tracks. Figure Ref \cite{47}.} 
\label{fig:10}
\end{figure}

Consisting of a stack of four silicon strips, the UT (located after the Velo) constitutes the second stage of track reconstruction. Due to the magnetic field in the UT region, the Velo tracks are extended to the UT based on a minimum momentum cut of 3 GeV (maximum bending allowed between Velo and UT) \cite{10}. This bending determines the track momentum. The UT hits are segregated into search windows by their x-coordinate and sorted by their y-coordinate, allowing for a quick comparison between the hits and the extended Velo tracks (see figure \ref{fig:10}). The best compatible hits (below a distance threshold) are appended to the existing Velo tracks \cite{47}. 

\setlength\belowcaptionskip{-3ex}

\subsubsection{Sci-Fi tracker}
\begin{figure}[h]
\includegraphics*[width=1\linewidth,clip]{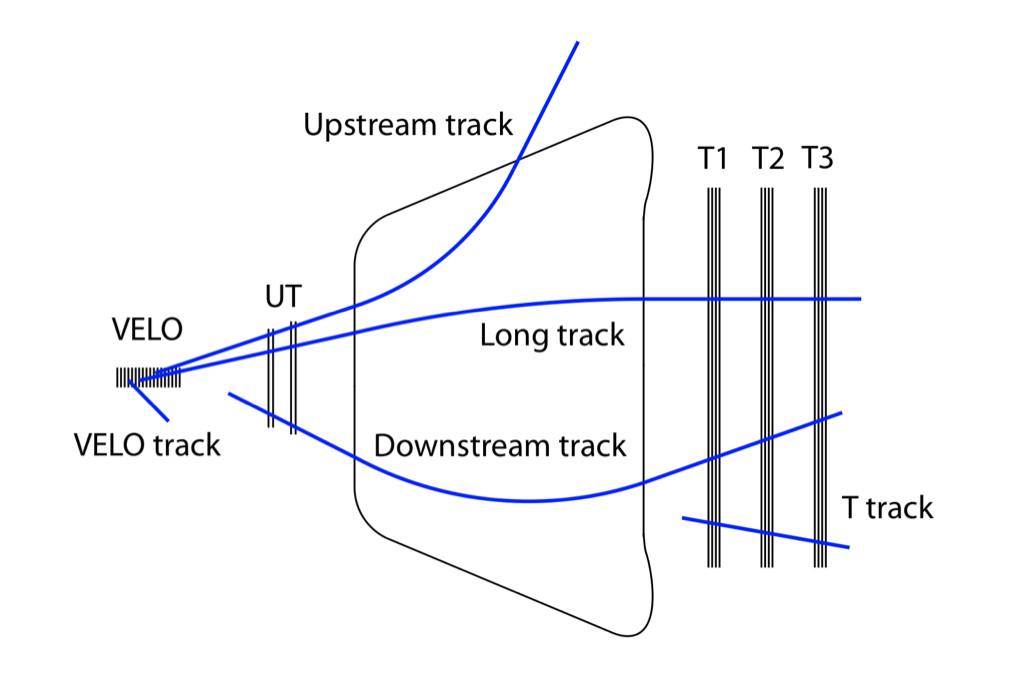}
\caption{Depiction of tracks transversing from the Velo \& UT detectors to the Sci-Fi detector (T1, T2, T3). Tracks are classified based on their direction. Low momentum particles are deflected the most by the B-field. Upstream tracks are reconstructed using the Velo \& UT hits. Long tracks don't always leave hits in the UT detector, hence are reconstructed using the Velo \& Sci-Fi hits. The downstream tracks are reconstructed using the UT \& Sci-Fi hits. Figure Ref \cite{25}.}\label{fig:11}
\end{figure}

The third stage of track reconstruction involves the Sci-Fi detector (placed after the magnet), consisting of three stations containing four scintillating fibres each. The four fibres within each station have a 50mm separation. These fibres are arranged in a stereo configuration (x-u-v-x), where the x fibres are aligned vertically, and the u and v fibres have a pitch of $\pm 5\degree$ respectively. The Velo-UT tracks are categorised based on their direction, which depends on their deflection by the magnet (see figure \ref{fig:11}).

\vspace{+5mm}

The Velo-UT tracks are extended through the Sci-Fi tracker by taking the track momentum and the magnetic field into account.  Loading the full magnetic field in memory is computationally expensive, which is why the magnetic field is parametrised as: 
\begin{align}
 B\textsubscript{x} \approx B\textsubscript{z} \approx  0
\end{align}
\vspace{-8mm}
\begin{align}
B\textsubscript{y}(z) = B\textsubscript{0}+B\textsubscript{1}\cdot z 
\end{align}

\vspace{-1mm}

Where $\frac{B\textsubscript{1}}{B\textsubscript{0}}$ is a constant \cite{10,48}. The Lorentz  equation is used to extend the Velo-UT tracks through the Sci-Fi stations:
\begin{equation} \label{eq:2}
\frac{\mathrm{d} \vec{p}}{\mathrm{d} t} = q\vec{v} \times \vec{B} 
\end{equation}

Track reconstruction at Sci-Fi uses the \emph{Hybrid seeding} algorithm, which generates track segments by seeding compatible Sci-Fi hits together \cite{49}. Since the hit efficiency of the fibres is about 98\% - 99\%, a minimum of five hits (in different x-layers) is required to classify a seed as a candidate track segment \cite{48}. The candidate segments with the lowest $\chi^{2}$ value relative to the extended Velo-UT tracks are attached to the Velo-UT tracks.

\subsubsection{Muon detector}
The final stage of track reconstruction occurs in the Muon detector, made up of four chambers separated by iron walls. Each chamber is segregated into four regions of different granularity, allowing the detector to deal with different amounts (rate) of particles passing through the detector \cite{50}. The scintillator pads measure the hits left by charged particles. The \emph{isMuon} algorithm \cite{51} deployed here extends the Velo-UT-Sci-Fi tracks to the muon detector and matches the hits from the muon chambers to the tracks using techniques similar to the UT stage. Ultimately, track properties such as the track momentum and the number of muon-chamber hits classify if a track corresponds to a muon or not.


\subsubsection{K\'alm\'an filter}
The second stage within track reconstruction after pattern recognition is the track fitting (estimating the track parameters). Conventional least-squares $\chi^{2}$ track fitting techniques require the availability of all measurements. Any new measurements (additional detector hits) requires a complete refit to be performed, which is time-consuming. Additionally, as charged particles travel through the various sub-detectors, they get deflected by multiple small-angle scattering due to their Coulomb interaction with the detector material, in turn altering the trajectory of the particle \cite{52}. HEP experiments depend on techniques like the Kalman filter (KF) to parametrise the tracks whilst considering the kinematical constraints and detector-material interactions. 

The Kalman filter (often referred to as a linear quadratic estimator (LQE) in statistics) is a technique which when applied to multiple noisy measurements, can estimate unknown variables (parameters) which describe the said measurements. It does this by constructing a joint probability distribution of the parameters spanning over multiple space or time separated intervals. This probabilistic approach allows the KF to produce more accurate estimates than estimates made from only a single measurement. The KF is applicable to fields as diverse as signal processing, robotics, guidance (Navigation) and rockets \cite{53}. 
In the context of HEP, the KF is applied to a track in three stages \cite{54}. The first stage projects the current state (track vector $\vec{r}_{i}$) to the next layer of the detector, described eq(10), following the same convention as \cite{55}.
\begin{equation}
\vec{r}_{i+1, proj} = F_{i} \vec{r}_{i} 	
\end{equation}
The covariant matrix $C_{i}$ of the current state is projected as:
\begin{equation}
C_{i+1, proj} = F_{i} C_{i}F^{T}_{i} + E_{i}
\end{equation}
Where $i$ is the index of the detector layers, $F_{i}$ is the transfer matrix (a linear operator which propagates the state), and $E_{i}$ is the error matrix which describes additive errors (\textbf{i.e} noise, multiple scattering effects). 

In the second stage, the projected state is corrected with the measurement (hit) belonging to this layer eq(12). This stage is skipped if a track has no associated hit in this layer. 
\begin{equation}
\vec{r}_{i+1, filt} = C_{i+1,filt} [C^{-1}_{i+1,proj}\vec{r}_{i+1, proj} + H^{T}G_{i+1}\vec{m}_{i+1}]
\end{equation}
Where
\begin{equation}
C_{i+1,filt} =  [C_{i+1,proj} + H^{T}G_{i+1}H]
\end{equation}
Where $H$ describes the relation between the projected state $\vec{r}_{proj}$ and the measurement   $\vec{m}$. The matrix $G_{i}$ describes the measurement noise (detector uncertainty). 
The final stage takes place once a track is forward fitted through all the detector layers. This stage involves back-propagating the tracks (improving the estimated track parameters using measurements/projections of the preceding layers).
Finally, the $\chi^{2}$ value eq(14) describes how well each track fits with its associated hits:
 \begin{equation}
 \chi^{2}_{t} =  \vec{R}^{T}_{i}G_{i} \vec{R}_{i} + (\vec{r}_{i,filt} - \vec{r}_{i,proj}) C^{-1}_{t,proj}(\vec{r}_{i,filt} - \vec{r}_{i,proj})	
 \end{equation}
Where $\vec{R}_{i}$ is the residue:
\begin{equation}
\vec{R}_{i} =  \vec{m} - H\vec{r}_{i, filt}
 \end{equation}
 
The KF is an inherently sequential technique. Since the pattern recognition stage occurs before the track fitting stage, the Kalman filter can fit individual tracks in parallel (hits belonging to different tracks don't need to be considered). The KF improves the impact parameter resolution of the tracks whilst outperforming the traditional global fit technique mentioned earlier. 
\vspace{-3mm}
\subsection{Vertexing}
\begin{figure}[H]
\includegraphics*[width=0.94\linewidth,clip]{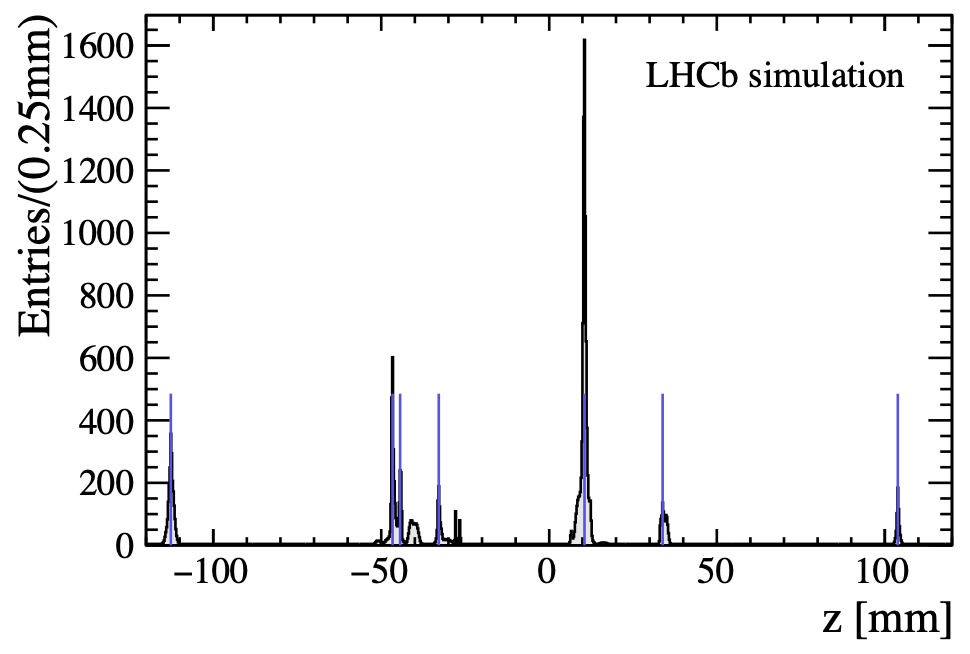}
\caption{Histogram describing the number density of the velo tracks (extrapolated back to the beamline) as a function of the z-position. The blue lines represent the peaks. The histogram is filled in parallel (each thread is associated to a track or a group of tracks). Fig Ref \cite{57}.} \label{fig:12}
\end{figure}

The second stage of event reconstruction is vertexing (clustering tracks that originate from the same point) which is crucial for making precise measurements. The enormous production of light quarks at the LHCb generates a significant background. Events containing $b\bar{b}$ pairs are differentiated by the observing decay vertices of b-hadrons displaced by a few millimetres from the corresponding primary vertex \cite{56}. Vertexing is crucial to discriminate between short and long-lived particles and is used as a part of the trigger (by looking for `displaced' signatures). The \emph{Fast parallel PV reconstruction} algorithm at the LHCb (proposed for Run 3) finds the PVs in four stages \cite{57}: 
\begin{itemize}[leftmargin=+0.4cm]
\itemsep-0.4em 
\item[(1)] The reconstructed Velo tracks are extrapolated backwards to the point of closest approach to beam-line.
\item[(2)] Since the z-axis is parallel to the beam line, a histogram of the density of tracks as a function of z-position is created. The peaks in this histogram correspond to primary vertex candidates (see figure \ref{fig:12}).
\item[(3)] The peaks in the histogram are identified.
\item[(4)] All tracks are partitioned to their respective PVs.  
\end{itemize}

\subsection{Particle identification}
Particle identification (PID) is a crucial part of event reconstruction. The physics goals of the LHCb experiment involves making precise measurements of hadronic final states. For example; two-body decays of $B^{0}$ and $B^{}_{s}$ to $\pi^{+}\pi^{-}$, $K^{+}\pi^{-}$ and $K^{+}K^{-}$ states \cite{50}. These final states have overlapping invariant mass peaks, which makes PID essential for distinguishing between these measurements. PID within the LHCb consists of three separate systems: the Ring Imaging Cherenkov (RICH) system (differentiates hadronic states), the calorimeter system (photon and electron identification, essential for radiative Penguin processes), and the Muon system (described earlier).

\subsubsection{RICH}
\begin{figure}
\includegraphics*[width=1.01\linewidth, height = 5.4cm]{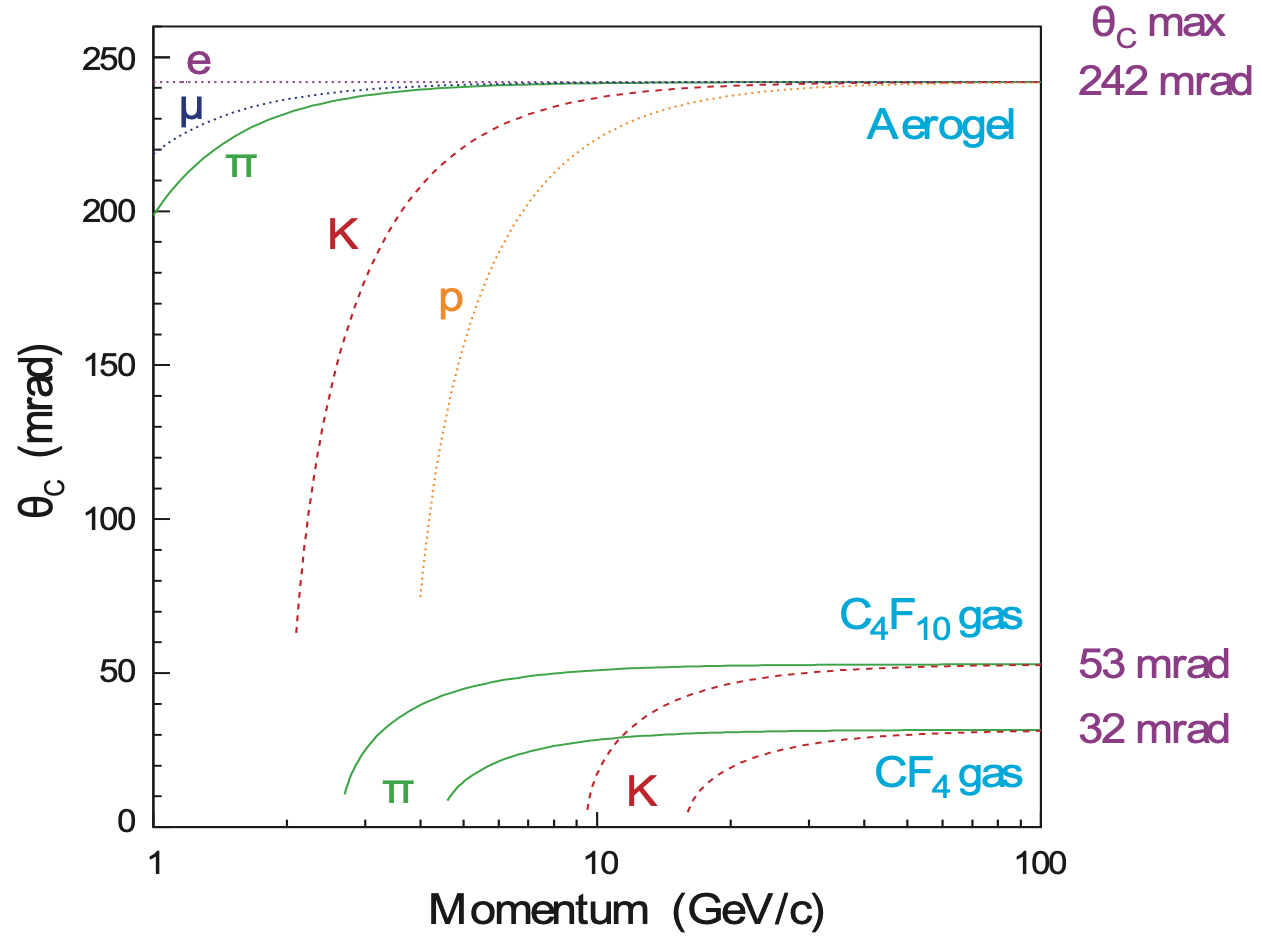}
\caption{A plot showing the distribution of the Cherenkov angle as a function of track momentum. The track momentum is obtained from the track fitting stage. Cherenkov angle is used to distinguish between different hadronic signatures. Fig Ref \cite{25}.} \label{fig:13}
\end{figure}

The LHCb contains two RICH systems (placed before and after the magnet). The RICH system depends on Cherenkov radiation \cite{7}: \textbf{i.e} EM radiation generated when a charged particle passes through a dielectric medium (silica Aerogel and $C_{4}F_{10}$ $CF_{4}$ gases in this case) faster than the phase velocity of the medium. Charged particles transversing through a medium radiate photons isotropically (forming Cherenkov rings), measured by the photodetectors. Once the track reconstruction stage is completed, the midpoint of the charged particle trajectory within RICH is assumed to be the emission point. The emission point is used to associate the measured photon hit with its respective parent track, in turn allowing to extrapolate the Cherenkov angle (emission angle) \cite{25}. The distribution of the Cherenkov angle as a function of the track momentum allows to distinguish between different hadronic states (see figure \ref{fig:13}). If the Cherenkov angle fits multiple hadron signatures, then each signature is assigned a weighted probability.

\subsubsection{Calorimeter}
\begin{figure}[h]
\includegraphics*[width=1\linewidth,clip]{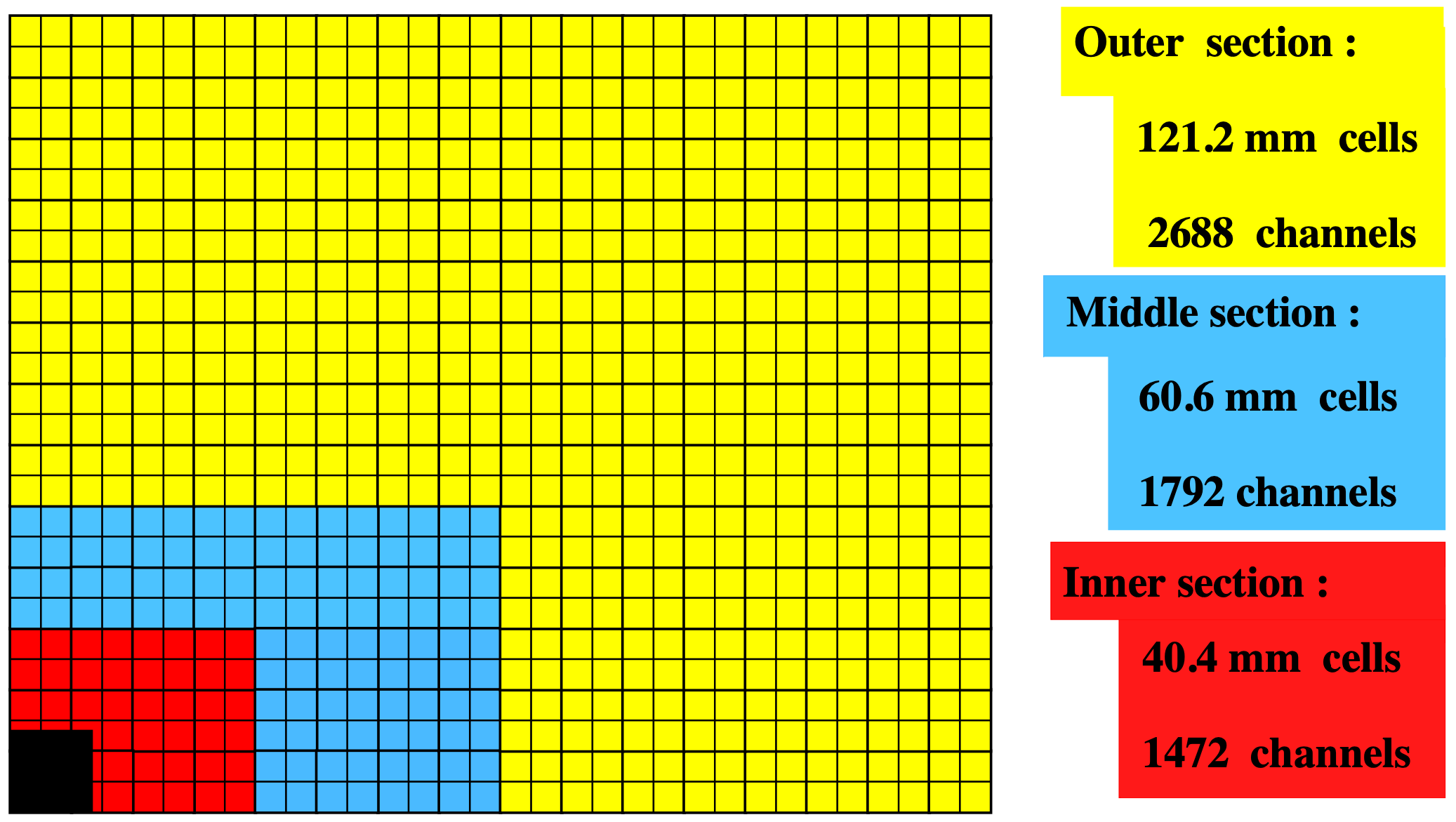}
\caption{Lateral segment of the ECAL module, representing a quarter of the detector's cross-section (\textbf{i.e}: where the bottom left corner corresponds to the centre of the detector). Fig Ref \cite{58}.} \label{fig:14}
\end{figure}

The LHCb contains two Calorimeters, the electromagnetic calorimeter (ECAL) followed by the hadronic calorimeter (HCAL), both placed after the magnet \cite{50}. This system provides the transverse energy of electrons, protons and hadron candidates. Both calorimeters operate on the same principle: scintillation light is transferred to the photomultiplier tubes (PMT) via the wavelength-shifting (WLS) fibres \cite{58}, in turn providing an energy measurement. Both ECAL and HCAL detectors follow a similar design. The cross-section (x-y plane) of the detector spans across the z-direction (parallel to the beamline), allowing particles to pass through it. The cross-section is divided into square channels/sensors (figure \ref{fig:14}). The hit density on the cross-section varies by two orders of magnitude \cite{58} (maximum hit density in the central region). For this reason, the central region of the detector has the highest density of channels/sensors. The ECAL and the HCAL modules have 5952 and 1468 channels each.   
\vspace{-3mm}

\section{Algorithms and Results}
\begin{figure}[H]
\includegraphics*[height = 4.7cm, width=1\linewidth]{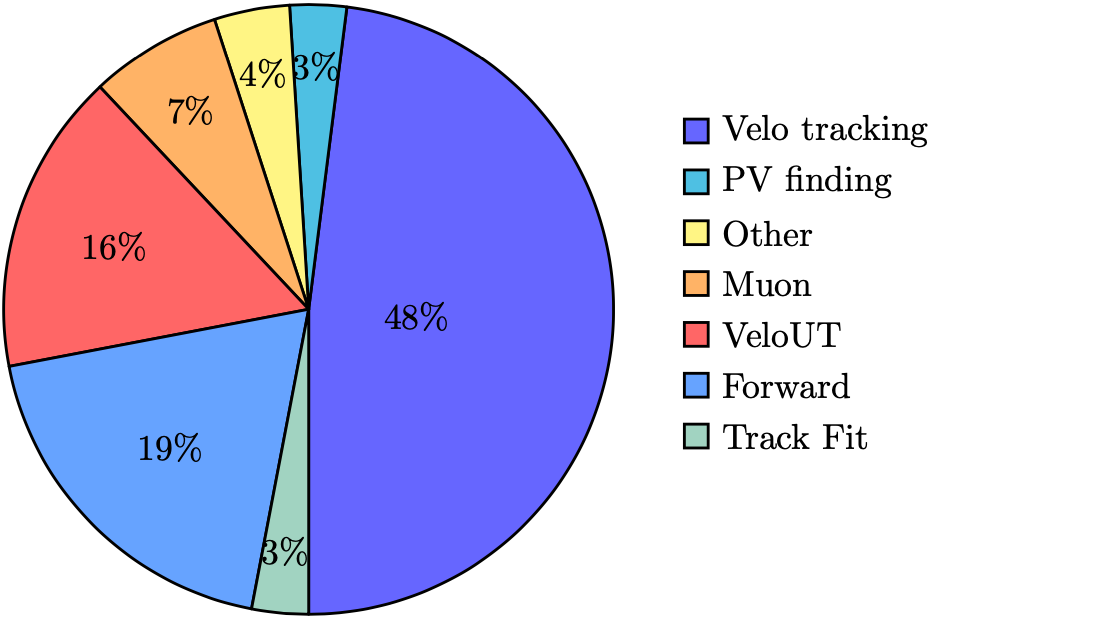}
\caption{Time fraction within the HLT1 sequence. Data Ref \cite{59}.} \label{fig:15}
\end{figure}

Within the HLT1 sequence, the Velo tracking stage consumes most of the computing resources (time). Figure \ref{fig:15} shows the time fraction of different processes within the HLT1 sequence. 
For this reason, three different Velo-track reconstruction algorithms (\textbf{i.e} predictive combinatorial seeding, artificial retina, a neural network model) and a vertexing algorithm were designed and implemented on the computing architectures discussed earlier in section 3. 
This section describes these algorithms' functionality, reconstruction performance, and scalability (computing time vs number of tracks on different computing architectures). Three metrics describe the reconstruction performance of an algorithm \cite{60}. Tracks are considered as \emph{reconstructible} if they at least leave three hits in different Velo modules.
\begin{itemize}[leftmargin=+0.4cm]
\itemsep-0.4em 
\item[(1)] The \emph{track reconstruction efficiency}: The probability of reconstructing a charged particle track, defined as the ratio of the number reconstructed (reconstructable) tracks over all the reconstructable tracks.
	\begin{equation}
	\frac{N_{reconstructed\;and\;reconstructable}}{N_{reconstructable}}
 	\end{equation}

\item[(2)] The \emph{fake track fraction}: The ratio between the reconstructed tracks which do not belong to any particles (fake tracks) over all reconstructed tracks.
	\begin{equation}
	\frac{N_{fake\;tracks}}{N_{reconstructed}}
 	\end{equation}

\item[(3)] The \emph{clone track fraction}: The ratio between the reconstructed tracks belonging to the same particle (duplicate tracks) over all reconstructed tracks.
	\begin{equation}
	\frac{N_{clone\;tracks}}{N_{reconstructed}}
 	\end{equation}
\end{itemize}
\vspace{+1mm}
An ideal track reconstruction algorithm would aim to maximise the track reconstruction efficiency whilst minimising the fake and clone track fractions. The simulated dataset of the Velo tracks and hits was supplied by Dr Daniel O\textquotesingle Hanlon. This dataset models the \emph{properties} of the Velo detector (\textbf{i.e} 26 silicon-strip modules on either side of the beamline, each module holding 14mm square sensors each with a resolution of 256 $\times$ 256 pixels). The hit efficiency of the modules (probability of recording a hit) is set to 98\%. For simplicity, the Velo-modules on either side of the beamline are aligned in z (same z-positions, unlike figure \ref{fig:8}). Every track reconstruction algorithm designed in this study obeys the following structure: due to the absence of a magnetic field, charged particles transverse the Velo detector in straight lines in 3-dimensions (no dependencies between the x and y coordinates of the hits). By exploiting this fact, the 3D track reconstruction problem is decomposed into two independent 2D track reconstruction problems, in the ($\hat{x}-\hat{z}$) and ($\hat{y}-\hat{z}$) planes. The track reconstruction algorithms here operate sequentially on each 2D plane. Once the 2D tracks in both planes are reconstructed, the \emph{shared} hits are used to couple the tracks from each 2D space to generate a 3D track. Results which contain/ignore multiple scatting effects are referred to as \emph{ms-data} and \emph{non-ms-data} respectively.

\subsection{The predictive combinatorial seeding algorithm}
This algorithm (designed from scratch) operates sequentially (locally)  from one module to the next, similar to the \emph{search by triplet} algorithm described in section 5.1.1. This algorithm functions in two stages.

\subsubsection{Design}
\vspace{-3mm}

\begin{figure}[H]
\includegraphics*[width=0.8\linewidth]{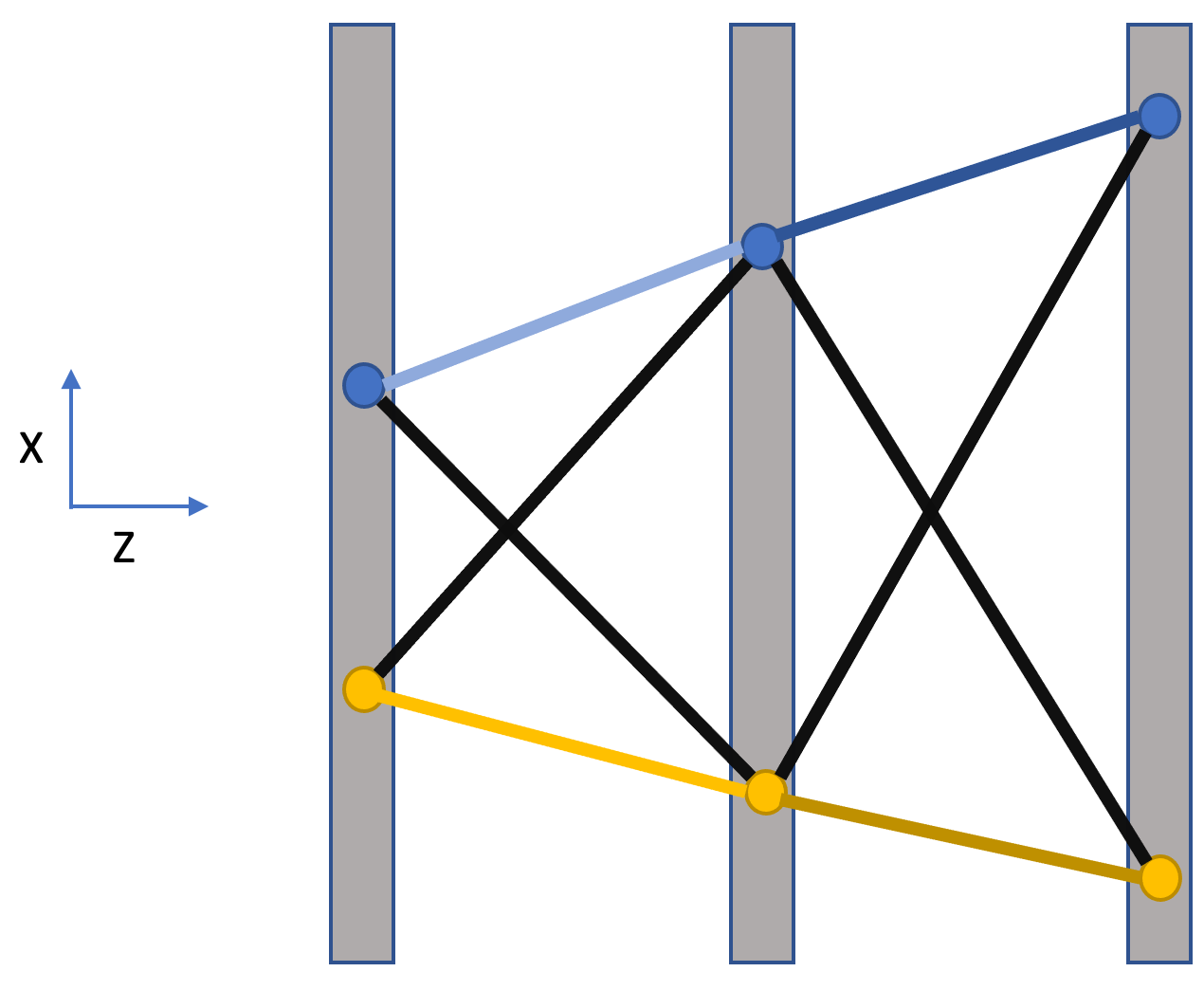}
\caption{A depiction of three  adjacent Velo modules, where hits belonging to the same track are colour coded. The lines represent all possible track segments between two adjacent detector modules.} \label{fig:16}
\end{figure}

1)\hspace{0.5 mm} Combinatorial stage: This stage operates on three subsequent detector modules at a time. Track parameters between the first-second module pair are calculated (\textbf{via} the equation of a line passing through two points) by taking all possible combination of hits between the first two modules. These parameters are stored in \emph{list-1}. The exact process is repeated for the hits present in the second-third module pair (stored in \emph{list-2}). The idea: line-segments in module$_{1\rightarrow2}$ and module$_{2\rightarrow3}$ belonging to the same track would produce the same track parameters. Hence list-1 and list-2 would share some elements (parameters) which define a possible Velo-track. Figure \ref{fig:16} illustrates this idea; here, three adjacent Velo-modules are shown. Two particles leave hits on these modules as they pass through them. Hits corresponding to the same track are colour coded (blue dots for track 1 and yellow for track 2). All combinations of track parameters are generated within modules$_{1\&2}$, creating the following line segments (light-blue, black, black, light-yellow). The same process is repeated in  modules$_{2\&3}$, creating (dark-blue, black, black, dark-yellow) line segments. The black line segments represent non-existent/invalid track parameters, and the light-dark line-segment variants of the same colour belong to the same track, which generate identical track parameters (stored in lists 1 and 2 respectively). Identical parameters are identified by comparing the two lists, and the common parameters are then appended to a global array for the next stage. In reality, the track parameters shared between the two lists would not be exactly identical due to multiple scattering effects and finite detector resolution. For this reason, a similarity threshold is used when comparing the elements of lists 1 and 2. The worst-case complexity of this stage is $\mathcal{O}(n^{2})$, where $n$ is the average number of hits in each module.

2)\hspace{0.5 mm} Prediction stage: This stage involves projecting the hits (defined by the track parameters calculated in the combinatorial stage) to the next Velo module. For example, the measured hits within the $4^{th}$ module are compared with the projected hits. A measured hit is \emph{flagged} if it overlaps with a projected hit. If all hits in the $4^{th}$ module are flagged, then the combinatorial stage (in the $4^{th}$ module) is skipped. If some hits remain unflagged (\textbf{i.e} new charged particle track originating between the $3^{rd}$ and the $4^{th}$ module), then the combinatorial stage is initiated. However, the flagged hits are ignored in this combinatorial stage since they already belong to a different track, reducing the workload drastically. The prediction stage operates sequentiality from one module to the next (flagging hits on the go), and the combinatorial stage is triggered when needed. Projecting and flagging the hits gives this stage a complexity of $\mathcal{O}(n \cdot log(n))$, where $n$ is the average number of hits in each module. Both stages heavily rely on control statements, (\textbf{e.g} deciding \textbf{if} a set of parameters are identical, flagging and appending a measured hit \textbf{if} it overlaps with a projected hit). Due to this combinatorial nature, the run time of this algorithm proliferates with the number of tracks. The worst-case complexity of the whole algorithm is $\mathcal{O}(n^{2})$, where $n$ is the average number of hits in each module.

\subsubsection{Results}
\begin{figure}[H]
\includegraphics*[width=0.97\linewidth]{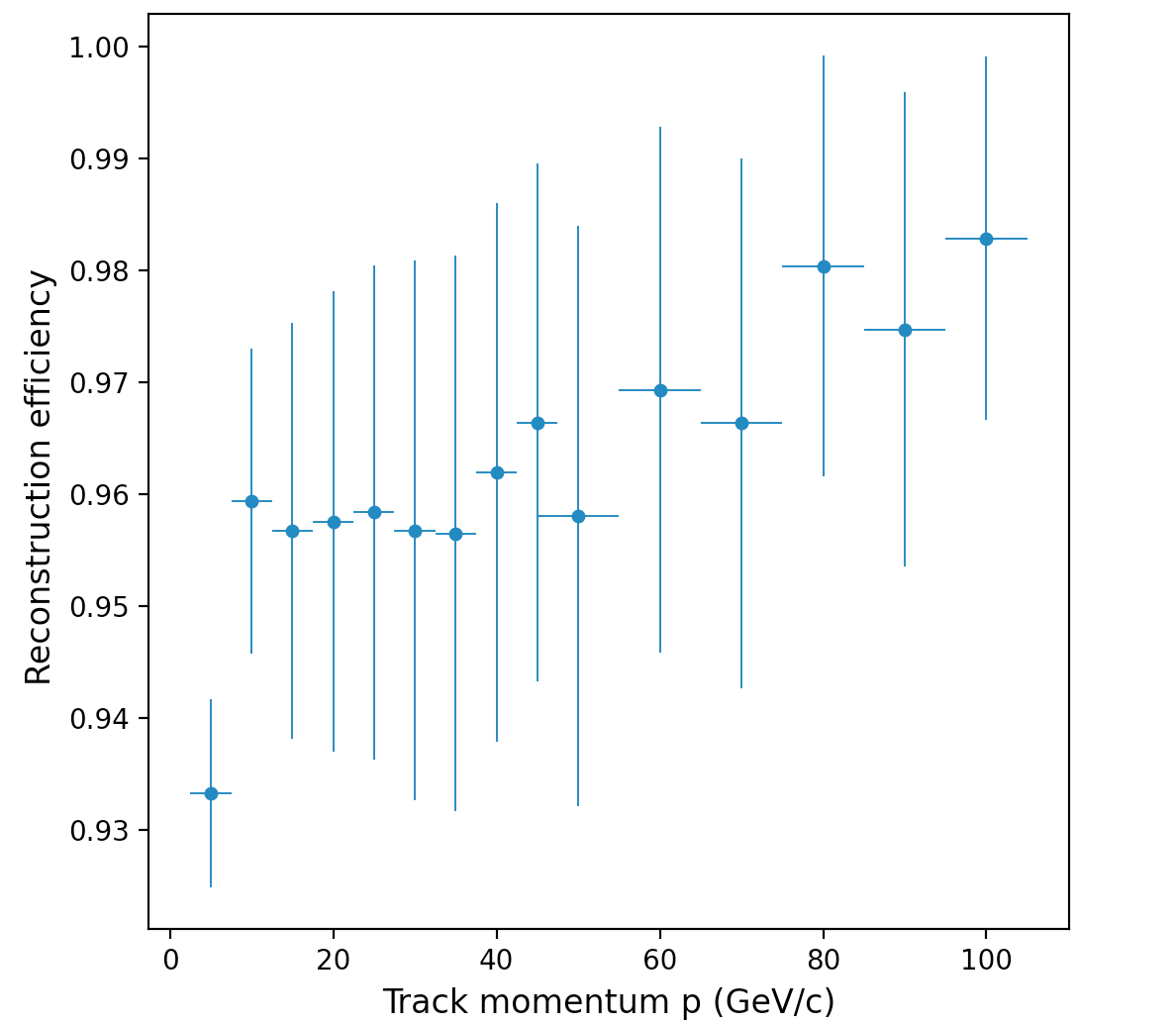}
\caption{Efficiency as a function of the track momentum (ms-data).} \label{fig:17}
\end{figure}

\begin{figure}[h]
\includegraphics*[width=0.97\linewidth]{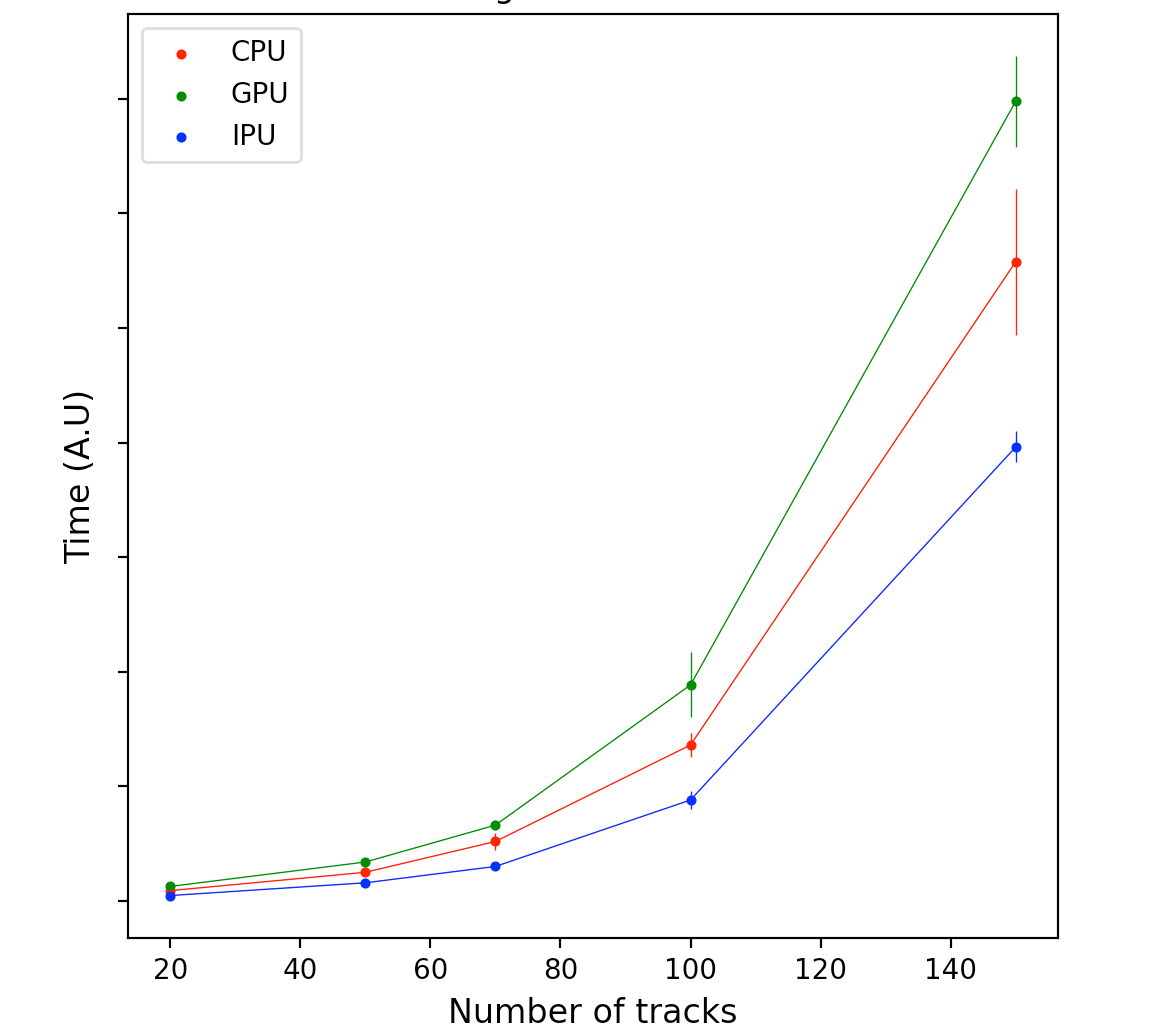}
\caption{Scaling of the computing time with the number of tracks, evaluated on the CPU, GPU, and the IPU. } \label{fig:18}
\end{figure}

\begin{figure}[H]
\includegraphics*[width=0.97\linewidth]{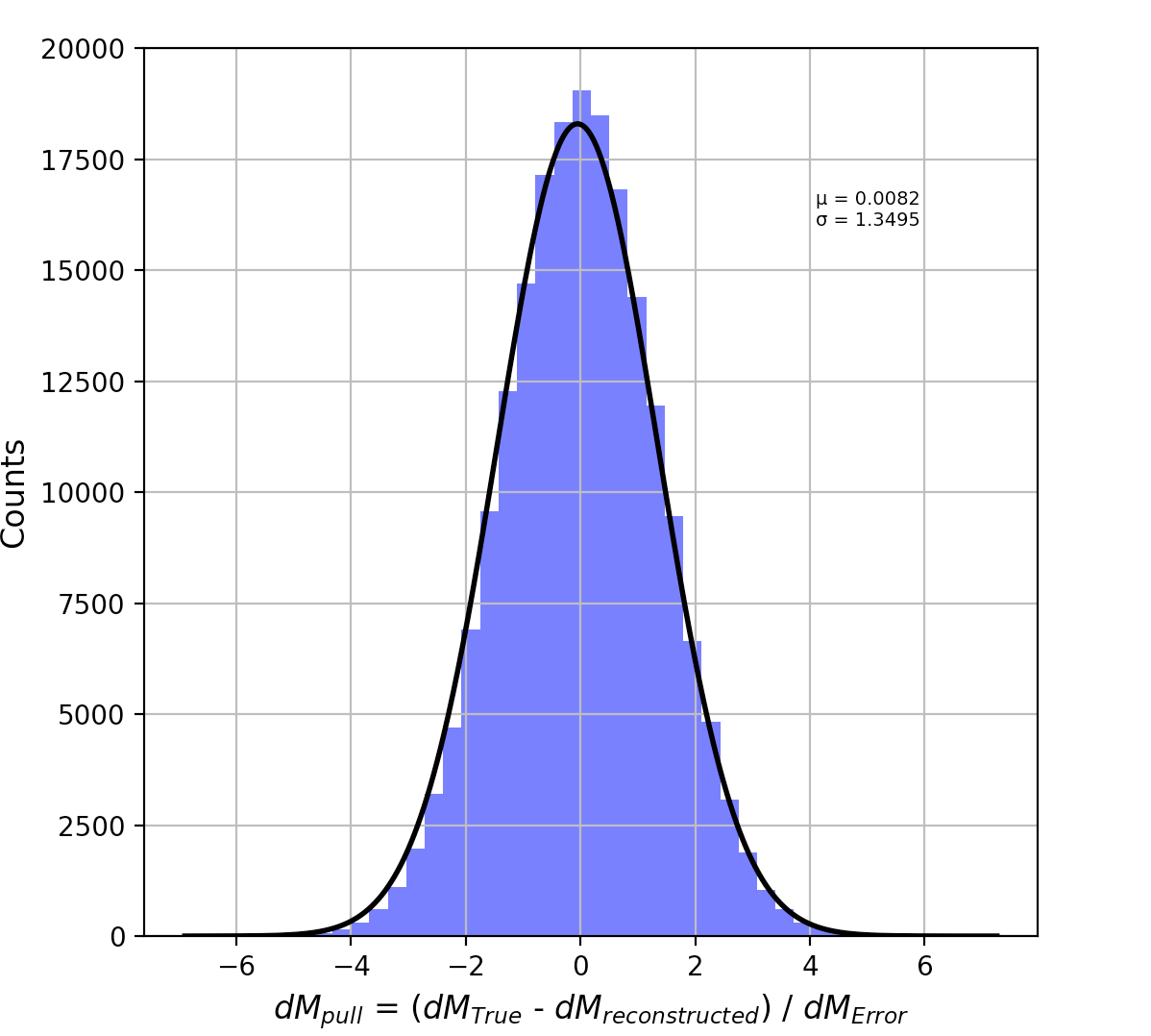}
\caption{Pull distribution of the reconstructed track parameters generated using 200000 simulated tracks (ms-data).} \label{fig:19}
\end{figure}

The reconstruction performance of this algorithm is summarised in table 2. Meanwhile, figures \ref{fig:17}, \ref{fig:18}, and \ref{fig:19} depict the efficiency measured against the track momentum,  the computing time as a function of the number of tracks, and the pull distribution of the track parameters respectively. The mean and the standard error of the measurements (efficiency, clone rate, fake rate, timing) were obtained by repeating the measurements on 2000 unique events, where each event contains about 200 reconstructible tracks. 

\begin{table}[h]
\begin{adjustbox}{width=\columnwidth,center}
\begin{tabular}{|c|c|c|c|c|c|}
\hline
\textit{Number of tracks} & \textit{10} & \textit{20} & \textit{50}                & \textit{80}                & \textit{100}               \\ \hline
\textit{Reconstruction efficiency}  & \textit{1}  & \textit{1}  & 0.972 $\pm$ 0.007 & \textit{0.967 $\pm$ 0.008} & \textit{0.958 $\pm$ 0.009} \\ \hline
\textit{Fake track fraction}      & \textit{0}  & \textit{0}  & \textit{0.035 $\pm$ 0.004}             & \textit{0.143 $\pm$ 0.005}             & \textit{0.281 $\pm$ 0.009}             \\ \hline
\textit{Clone track fraction}     & \textit{0}  & \textit{0}  & \textit{0.037 $\pm$ 0.005}             & \textit{0.086 $\pm$ 0.006}             & \textit{0.133 $\pm$ 0.008}             \\ \hline
\end{tabular}
\end{adjustbox}

\caption{Track reconstruction performance of the predictive combinatorial seeding algorithm as a function of the number of tracks (non-ms-data).}
\label{tab:my-table}
\end{table}

As the number of tracks increases from 10 to 100, the reconstruction efficiency decreases by 4\%; meanwhile, the fake and clone track fractions increase to 28\% and 13\% (table 2). The reconstruction efficiency decreases because, with an increasing number of reconstructable tracks, the probability of multiple tracks being extremely close increases (small angular separation resulting in track parameters being close in value). If the difference between these track parameters is below the threshold described in the combinatorial stage, the algorithm would be unable to resolve such tracks (considering them as one). Additionally, since only three hits are required to define a candidate Velo-track, the increasing hit density in each module (proportional to the number of tracks) makes it more likely for hits belonging to different tracks in different modules to line up (forming a fake/clone candidate), which explains the increase in the fake and clone track fractions.

\setlength\belowcaptionskip{-0.9ex}

The multiple scattering deflection (angle) is inversely proportional to the track momentum \cite{61}. Low momentum tracks ($<$ 10 GeV) are strongly deflected by the Coulomb interaction with the detector material. Hence, the reconstruction efficiency improves with increasing track momentum (figure \ref{fig:17}).

In terms of computing performance (figure \ref{fig:18}), the IPU outperforms the CPU, which surpasses the GPU. The average speedup ratios (CPU/GPU and CPU/IPU) are 0.7 and 1.7 respectively. The poor performance of the GPU can be explained by assuming that the TensorFlow's XLA (Accelerated Linear Algebra) backend is mapped on the GPU hardware such that it can utilise parallel threads/blocks. If such is the case, then diverging workflows due to control flow statements (\textbf{i.e} parallel threads identifying and appending a different number of track parameters due to the threshold conditions) results in load imbalance. This generates significant performance penalties since the SIMT architecture of the GPU succumbs to the effects of warp divergence \cite{62}. In contrast, the IPU is immune to irregular/diverging workflows due to its MIMD nature, allowing each tile to operate independently.

\subsection{The artificial retina algorithm}
The artificial retina algorithm is a global algorithm that estimates the track parameters by operating on all hits at the same time, unlike the sequential algorithms described earlier (\textbf{i.e} \emph{search by triplet} and \emph{predictive combinatorial seeding}). First proposed by \emph{L. Ristori} in 2000 \cite{63}, the artificial retina algorithm takes inspiration from the visual cortex of mammals. The mammalian retina responds to a stimulus (light) by sending chemical and electrical signals to the visual cortex, which assimilates this signal to produce an image. The mammalian retina consists of millions of neurons, and each neuron is calibrated to recognise a particular shape (\textbf{i.e} properties such as edges and orientation) on a specific region of the retina known as a receptive field \cite{64}. The similarity between the shape of the stimulus and the shape the neuron is calibrated to recognise determines the intensity (strength) of the neuron's response. Hence different neurons react to a common stimulus with varying intensities (weights). The visual cortex extrapolates the structure of the stimulus by using these weights.  
The artificial retina algorithm borrows this notion of a `neuron's response' and applies it to the discipline of track reconstruction in HEP.

\vspace{-2mm}
\subsubsection{Design} 
\begin{figure}
\includegraphics*[width=1\linewidth]{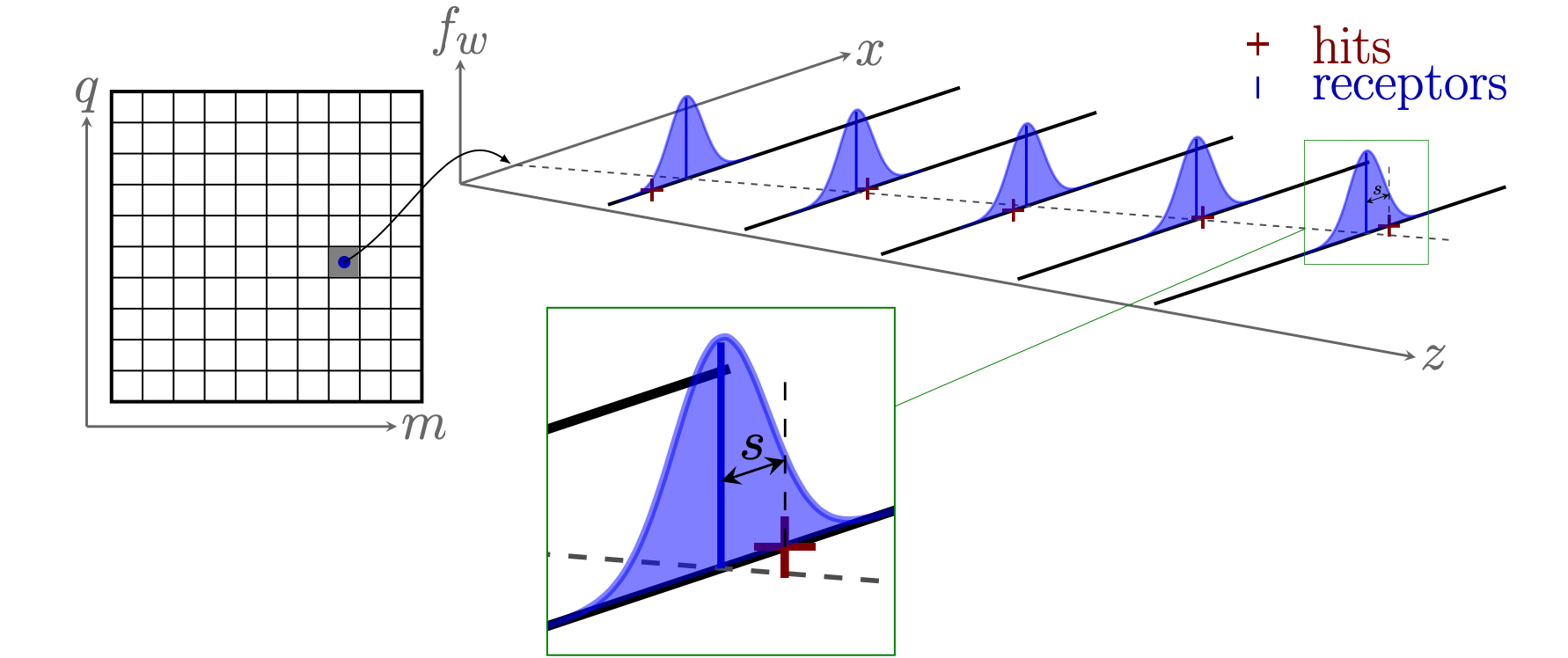}
\caption{A representation of the discrete two dimensional ($q,m$) parameter space on the left. A single neuron/track parameter ($q_{i},m_{j}$) is used to draw a hypothetical track. The receptors are defined as the points of intersection of this track with the detector layers. The distance ($s$) between a receptor and a measured hit determines the contribution to the overall intensity value of a cell. (\textbf{i.e} smaller the distance, larger the contribution). Fig Ref \cite{65}.} \label{fig:20}
\end{figure}

The core idea behind the artificial retina algorithm is that each (2D) Velo-track is described by two parameters ($q,m$).
A $200 \times 200$ grid is created within the two dimensional parameter space ($q,m$), where each neuron within the grid ($q_{i},m_{j}$) defines a particular track, giving a total of $200^{2} = 40000$ neurons/track parameters.
The intersection point of a track ($q_{i},m_{j}$) with the detector modules (hereinafter referred to as a receptor) $v^{k}_{i,j}$ is calculated, where $k$ is the Velo-module index. An intensity value for each neuron within the grid is calculated via eq(19).
\begin{equation}
I_{ij} = \sum_{k,r}^{} exp(\frac{-s_{ijkr}^{2}}{2\sigma })	
\end{equation}
Where
\begin{equation}
s_{ijkr} = v_{r}^{k} - v^{k}_{i,j}	
\end{equation}
is the distance between one measured hit $v_{r}^{k}$ and the receptor $v^{k}_{i,j}$ in the $k^{th}$ module. This sum is extended over all hits in every detector module. The value of $\sigma$ can be tuned to improve the sensitivity/response of the receptors \cite{63}. For a single term within the sum in eq(19), the smaller the distance ($s_{ijkr}$) between a hit and a receptor, the stronger the response of a receptor; this is illustrated in figure \ref{fig:20}.

A loop iterates over the 40000 neurons in the grid, calculating their individual intensities. Track parameters (neurons) that fit the data well produce high-intensity responses. If a neuron generates a high-intensity response, the chances are, its neighbouring neurons (\textbf{i.e} close in value) may generate high/moderate intensity responses as well, creating a cluster of `excited' neurons. 
Once the intensities have been calculated, neurons above a certain intensity threshold are shortlisted. The hierarchical clustering package \emph{scipy.cluster.hierarchy} \cite{66} is used to further shortlist single/clusters of neurons. A neuron cluster is reduced to a single neuron by calculating a weighted average of the central neuron and its neighbours. The neurons which survive these two filtering stages are used to generate the Velo-tracks. This algorithm (excluding the scipy filtering stage) scales as $\mathcal{O}(n)$ where $n$ is the total number of Velo-hits.

\subsubsection{Results} 
\begin{figure}[H]
\includegraphics*[width=0.97\linewidth]{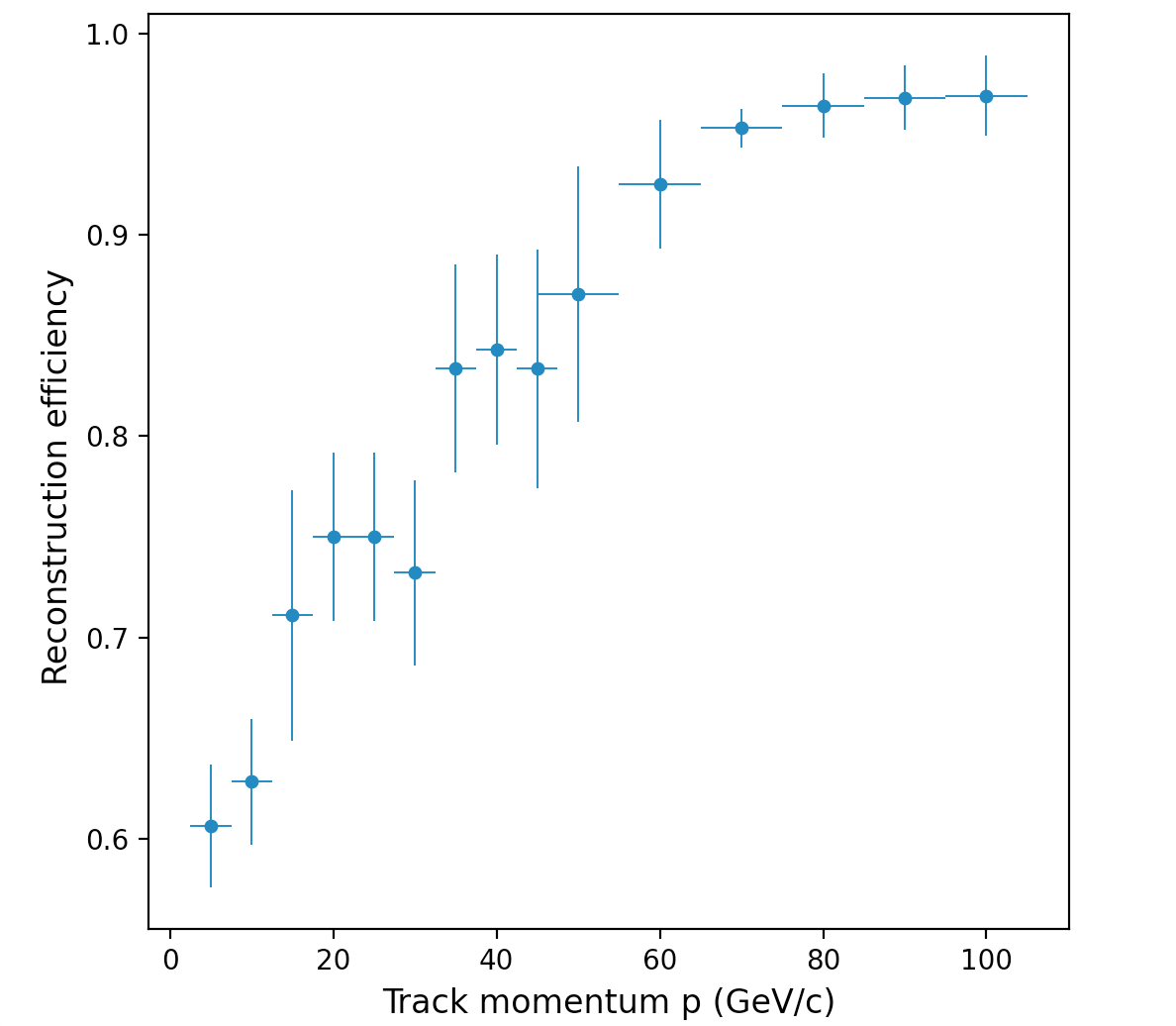}
\caption{Efficiency as a function of the track momentum (ms-data).} \label{fig:21}
\end{figure}

\begin{figure}[H]
\includegraphics*[width=0.97\linewidth]{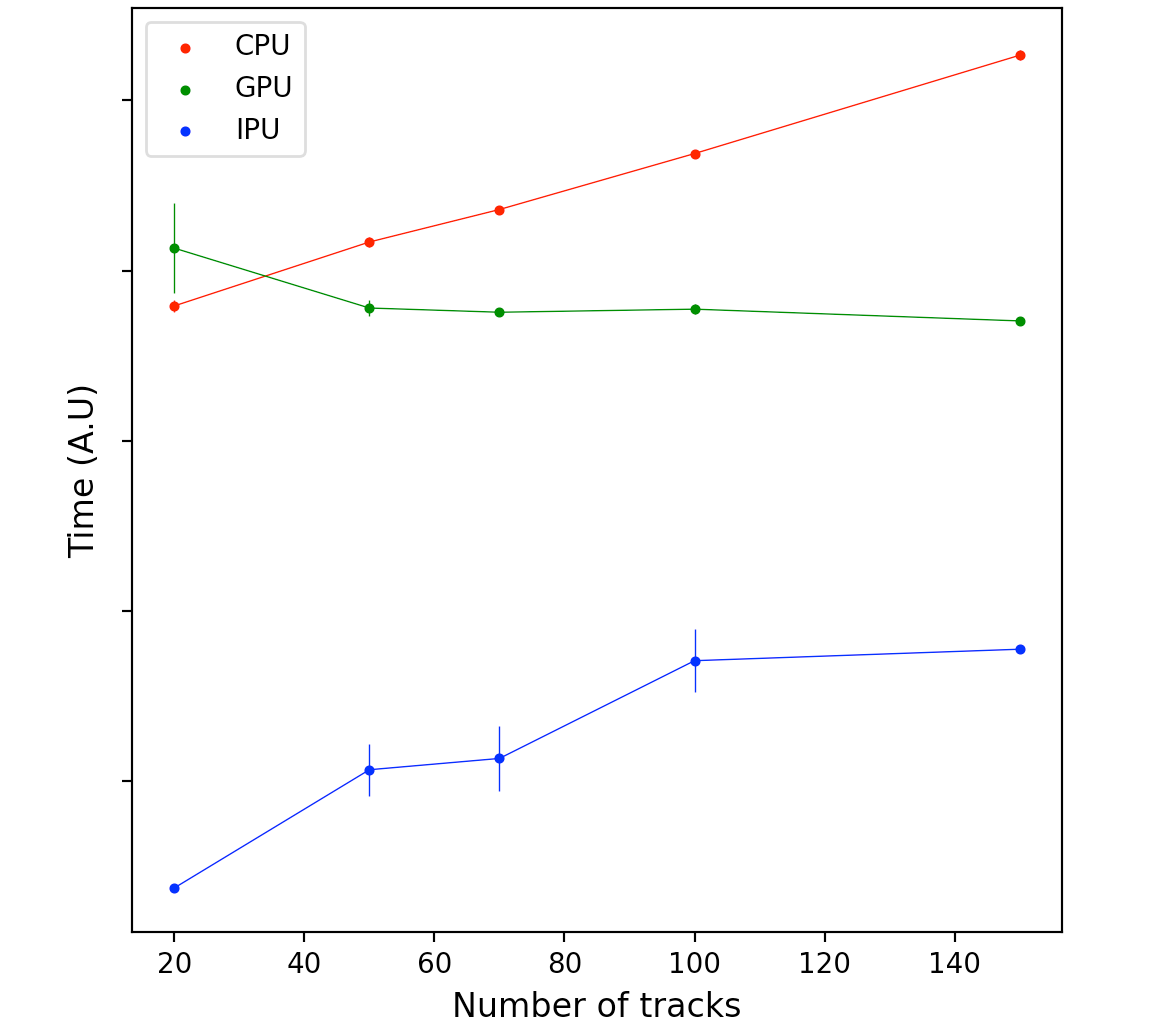}
\caption{Scaling of the computing time with the number of tracks, evaluated on the CPU, GPU, and the IPU.} \label{fig:22}
\end{figure}

\begin{figure}[H]
\includegraphics*[width=0.97\linewidth]{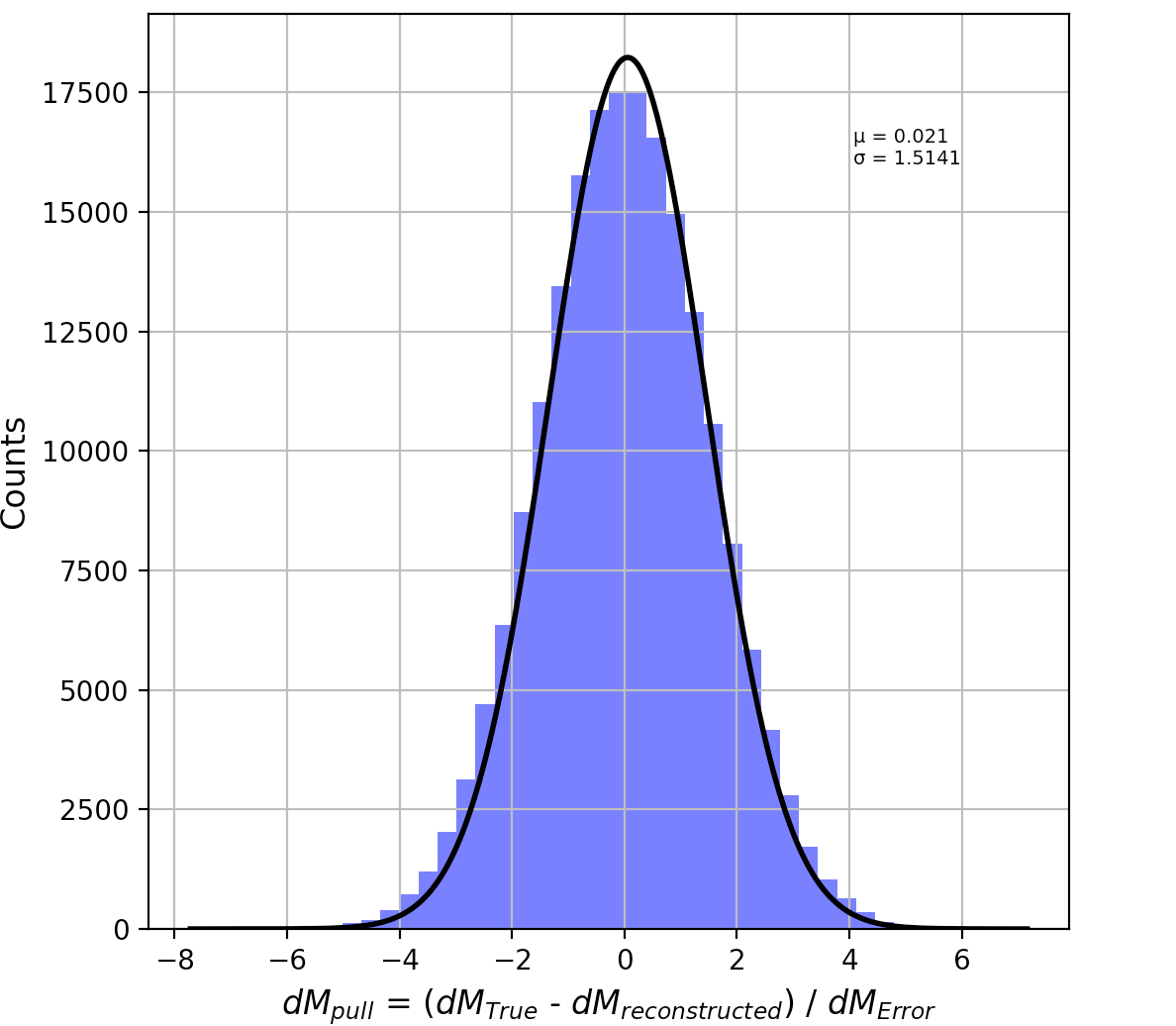}
\caption{Pull distribution of the reconstructed track parameters generated using 200000 simulated tracks (ms-data).} \label{fig:23}
\end{figure}

The reconstruction performance of the artificial retina algorithm is summarised in table 3. Figures \ref{fig:21}, \ref{fig:22}, and \ref{fig:23} depict the efficiency measured against the track momentum, the computing time as a function of the number of tracks, and the pull distribution of the track parameters respectively.
\begin{table}[h]
\begin{adjustbox}{width=\columnwidth,center}
\begin{tabular}{|c|c|c|c|c|c|}
\hline
\textit{Number of tracks} & \textit{10} & \textit{20} & \textit{50}                & \textit{80}                & \textit{100}               \\ \hline
\textit{Reconstruction efficiency}  & \textit{0.925 $\pm$ 0.004}  & \textit{0.923 $\pm$ 0.004}  & 0.896 $\pm$ 0.006 & \textit{0.863 $\pm$ 0.007} & \textit{0.844 $\pm$ 0.009} \\ \hline
\textit{Fake track fraction}      & \textit{0.065 $\pm$ 0.001}  & \textit{0.072 $\pm$ 0.001}  & \textit{0.109 $\pm$ 0.003}             & \textit{0.174 $\pm$ 0.002}             & \textit{0.232 $\pm$ 0.002}             \\ \hline
\textit{Clone track fraction}     & \textit{0.186 $\pm$ 0.003}  & \textit{0.232 $\pm$ 0.003}  & \textit{0.251 $\pm$ 0.007}             & \textit{0.291 $\pm$ 0.006}             & \textit{0.298 $\pm$ 0.008}             \\ \hline
\end{tabular}
\end{adjustbox}

\caption{Track reconstruction performance of the artificial retina algorithm as a function of the number of tracks. (non-ms-data)}
\label{tab:my-table}
\end{table}

Like the combinatorial algorithm, the mean and the standard error were obtained by repeating the measurements over 2000 events. The reconstruction efficiency of the retina algorithm drops to 84\% as the numbers of tracks increase from 10 to 100, whereas the fake and clone track fractions increase to 23\% and 29\% respectively (table 3). The drop in reconstruction efficiency is almost four times that of the combinatorial algorithm. The reconstruction efficiency decreases because the resolution of the reconstructed track parameters here is quite limited compared to the combinatorial algorithm. For example, if two or more reconstructable tracks are extremely close together such that they are all associated with a single neuron or neighbouring neurons, then the retina algorithm would be unable to distinguish between such tracks.
The resolution of the neuron/track parameter depends on the number of discrete neurons the grid is divided into (\textbf{i.e} the bin size). Increasing the bin size would improve the reconstruction performance at the expense of extra computation since this algorithm scales quadratically $\mathcal{O}(m^{2})$ with the bin size $m$.

The fake track fraction of the retina algorithm is comparable to the combinatorial algorithm; meanwhile, the retina algorithm generates twice as many clone tracks. This is primarily because of the excitation of multiple neurons by a set of hits belonging to a common track. These neurons are close to each other but may not necessarily be neighbours (\textbf{i.e} near neighbouring neurons or neighbours of neighbours). The scipy clustering stage only looks for neighbouring neurons, which explains why these (clone/similar) neurons are overlooked and are considered as distinct tracks.

The reconstruction efficiency improves with the track momentum for the same reasons as explained in section 6.1.2 (figure \ref{fig:21}). The artificial retina algorithm is a lot more sensitive to multiple scattering compared to the combinatorial algorithm. This is primarily because the small-angle scattering changes the distance between the receptors and the measured hits. The intensity function eq(19) is a sensitive function of this distance (\textbf{i.e} large distance/deflection reduces the intensity contribution of the receptors), making it more likely that a `correct' neuron is filtered out. 
Increasing the value of $\sigma$ in eq(19) improves the reconstruction performance for low momentum tracks at a loss of the receptor's sharpness (\textbf{i.e} granularity with which it can resolve different hits), which increasing the fake and clone track fractions. The value of sigma was set to $1/N_{bins} = 1/200$ across all measurements.

In terms of computing performance, both the GPU and the IPU outperform the CPU (figure \ref{fig:22}). The average speedup ratios (CPU/GPU and CPU/IPU) are 1.2 and 1.6 respectively. This is mainly because calculating the intensity of each neuron by iterating over the parameter grid is an inherently parallel process (\textbf{i.e} each iteration is identical and independent). The results can be explained by again assuming that the XLA backend is mapped on the GPU and the IPU such that the intensity calculation is distributed across multiple cores, whereas the CPU performs this calculation sequentially. Furthermore, unlike the combinatorial algorithm, the retina algorithm involves no branches (conditional statements). This regular/uniform computation maps well on the GPU.

\subsection{Convolutional neural network based track reconstruction}
\subsubsection{Supervised Machine Learning basics}
Supervised machine learning (ML) problems typically involve searching for some unknown function  $f_{\phi} : \textbf{X} \rightarrow \textbf{Y}$, where \textbf{X} and \textbf{Y} are the inputs (observed data), and the target outputs respectively, here $\phi$ represents the parameters of the function $f_{\phi}$. Such a function could either map the input to a continuous value (regression) or it could classify the input into a predefined set of classes (classification). A metric known as the loss function $L(\textbf{y},f_{\phi}(\textbf{x}))$ describes the difference between the predicted output $f(\textbf{x})$ and the desired output $\textbf{y}$ \cite{67}. Two common regression loss functions are:
\vspace{-2mm}
\begin{align}
 \ \text{Mean Square Error (MSE)} = \frac{1}{N} \sum_{i = 1}^{N} (y_{i} - f_{\phi}(\textbf{x}_{i}))^{2} \\
   \text{Mean Absolute Error (MAE)} = \frac{1}{N} \sum_{i = 1}^{N}|y_{i} - f_{\phi}(\textbf{x}_{i})| 
 \end{align}
 By using a training dataset \{$\textbf{x}_{i},\textbf{y}_{i}\}^{N}_{i = 1}$ (set of inputs and their target outputs), supervised ML algorithms such as: neural networks, decision trees, support vector machine try to minimise the loss function by tuning the parameters ($\phi$) of the unknown function $f_{\phi}(\textbf{x})$.
The goal of a supervised ML algorithm (after minimising the loss function) is to make accurate predictions (outputs) using data (inputs) that were not used to train the model \cite{68}.

\subsubsection{Neural Networks}

Neural Networks (NN) are a subset of supervised machine learning techniques used for extrapolating the underlying rela-
\begin{figure}[H]
\includegraphics*[width=8cm, height = 5.7cm]{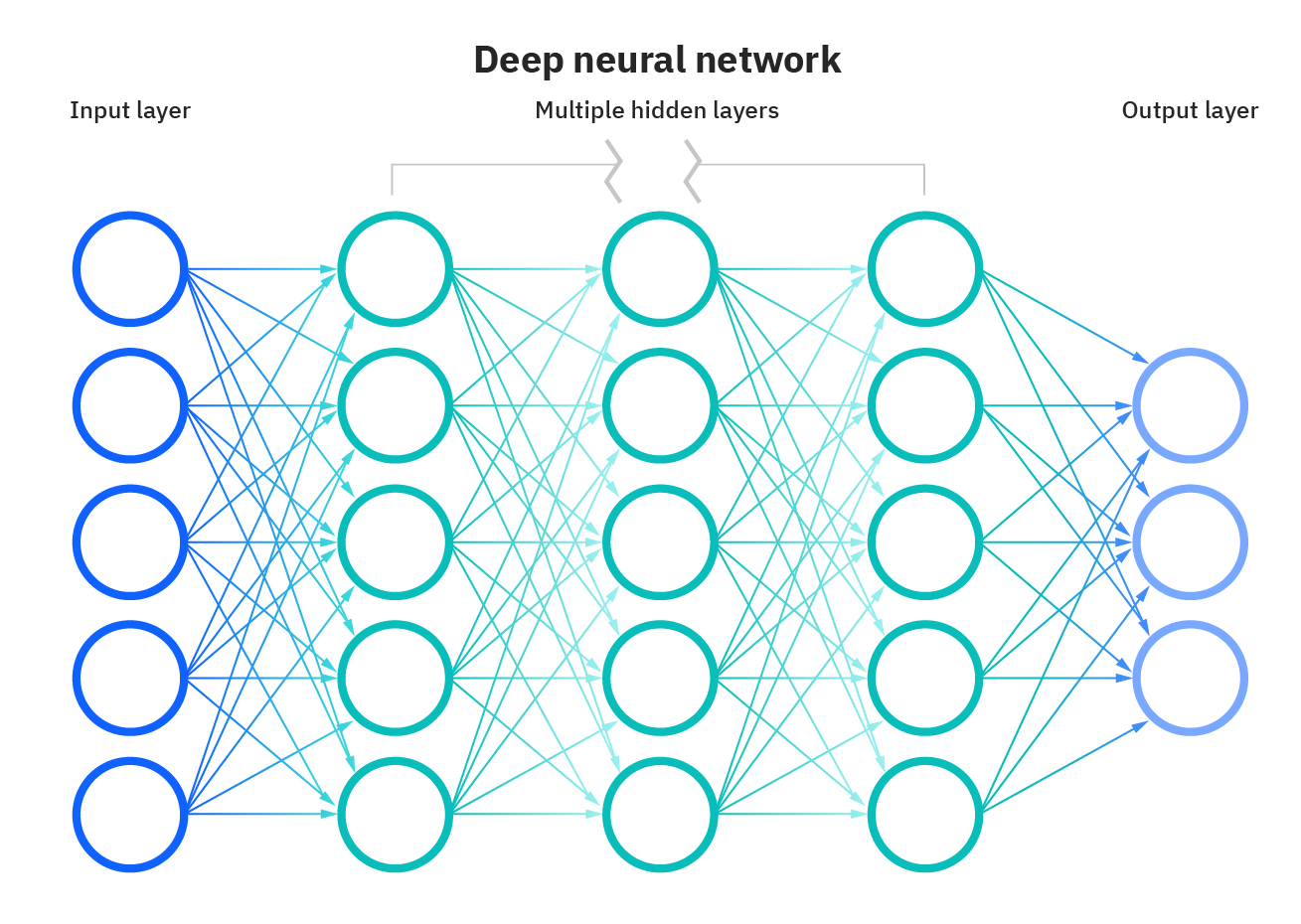}
\caption{Depiction of a deep neural network with three hidden layers. This network represents a series of transformations starting from an input vector of size 5 to an output vector of size 3. The transformation from one layer to the next is described by a matrix multiplication, as shown in eq (23). Image Ref \cite{69} } \label{fig:24}
\end{figure} 
\begin{figure*}
\captionsetup{singlelinecheck = false, justification=justified}
  \includegraphics[width=1\linewidth,clip]{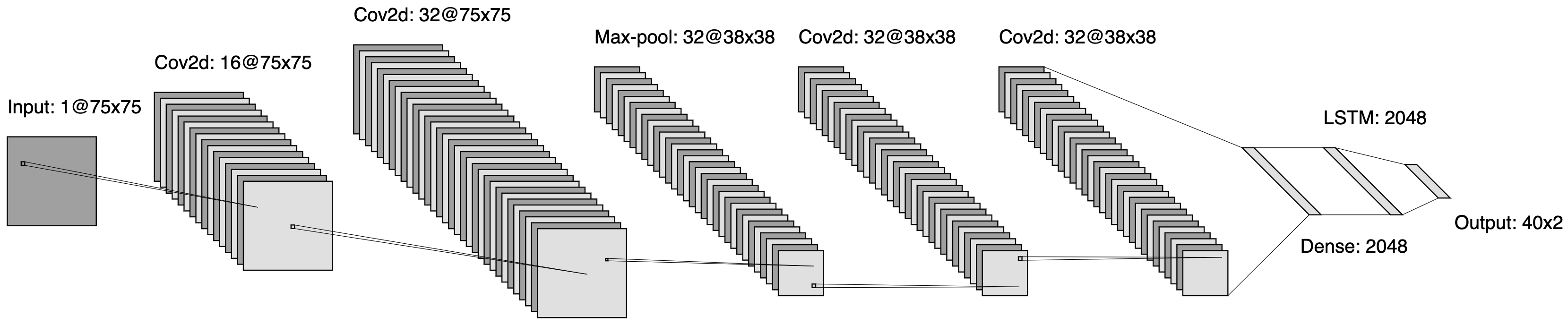}
  \caption{Neural network architecture containing CNN and RNN (LSTM) layers for velo-track reconstruction. } \label{fig:25}
\end{figure*}
-tionship between the input and the output states by a process that mimics the way the human brain processes information. 
A neural network encompasses a series of transformations from an initial state (input $\textbf{x}$) to a final state (output $\textbf{y}$) with few (shallow NN) or many (deep-NN) intermediate states known as `hidden' states $\textbf{h}_{i}$. The structure (size) of a neural network defines the space of functions in which the unknown function $f_{\phi}$ can be searched. Each transformation is represented by a vertical layer of `neurons', as seen in figure \ref{fig:24}. The following equation describes the transformation from one layer to the next layer:
\begin{equation}
\textbf{h}_{i+1} = g_{i}(W_{i}\textbf{h}_{i} + b_{i})	
\end{equation}

Where, the subscript $i$ denotes the layer number, $\textbf{h}_{i+1}$ is the hidden state in the next layer, $g_{i}$ is the activation function, $\textbf{h}_{i}$ is the state in the current layer, and $b_{i}$ is a vector bias. This linear transformation is governed by the matrix $W_{i}$ whose elements are referred to as weights. A NN is described by applying such transformations sequentiality from one layer to the next. Initially, the weights and the biases are randomised, and then the NN is `trained', which reduces the loss function. This is done in two stages. The first stage, known as back-propagation involves calculating the gradient of the loss function with respect to the weights $\nabla_{\phi} L(\textbf{y},f_{\phi}(\textbf{x}))$. The second stage, known as gradient descent, exploits the fact that the negative of each gradient calculated in the back-propagation stage points in the direction which reduces the loss function, allowing it to update the weights to do so. A NN is trained on the training samples, and a cycle of training through all the samples is known as an epoch. Typically a NN is trained over $\mathcal{O}(10)$ epochs.

Neural networks are categorised based on their composition of modular differentiable components \cite{67}. The convolutional neural network (CNN) is one such example and is arguably the most critical network in image analysis.  The CNN operates on a shared weight architecture of convolution filters, which allows it to identify specific features (\textbf{i.e} edges, brightness, sharpness, contrast) regardless of where these features appear in an image (\textbf{i.e} translation equivariant responses) \cite{70}. Another important neural network architecture is the recurrent neural network (RNN). Unlike the feed-forward architecture of the DNNs and CNNs, where the input state is transformed at once, the RNN consumes the input state in a sequence (one at a time). At each sequence, the RNN operates on a part of the input, producing a hidden output. This output is forwarded to the following sequence of calculations, and this continues till the entire input is consumed \cite{71}. This makes the RNN well suited for analysing sequential (temporal) data such as speech recognition, time series prediction, and conversational interfaces such as chatbots \cite{72}.

\subsubsection{Design}
The approach here involves a neural network that uses convolution and recurrent layers to estimate the Velo-track parameters. Unlike the algorithms described earlier, which process hits (position data) to find the tracks, this approach estimates the track parameters visually by processing a $75 \times 75$ pixel image of the hits within the Velo-space. Straight tracks within the pixel space are generated by a simple toy dataset, ignoring any multiple scattering effects for simplicity. This pixel space contains 26 modules that record the hits (mimicking the Velo setup). The z-spacing between the modules is consistent with the Velo setup described in section 6.0. The pixels in the Velo-modules which intersect the tracks are set to 1 and all other pixels are set to 0.
\begin{figure}[H]
    \hbox{\hspace{+1.9em}\includegraphics*[width=0.8\linewidth]{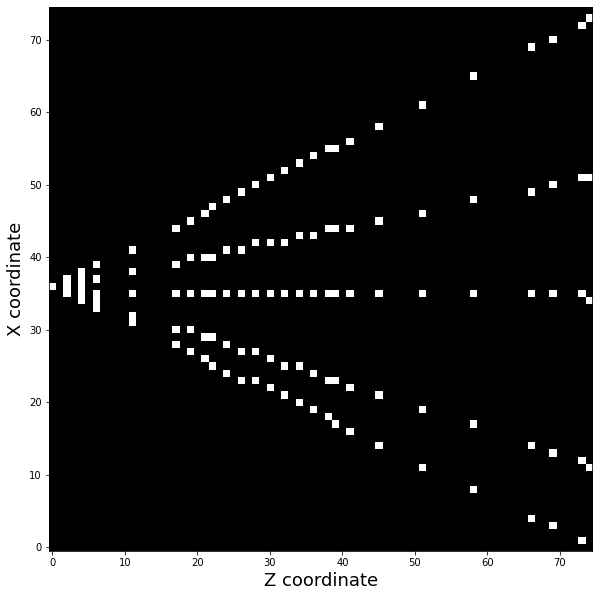}}
    \caption{Figure depicting five straight tracks in the ($75 \times 75$) velo-pixel space. This space contains a total of 26 velo-modules. } \label{fig:26}
\end{figure} 
Figures \ref{fig:25} and \ref{fig:26} represent the architecture of this neural network and five tracks in the Velo-pixel space respectively. 
This network was trained on 320000 training samples ($75 \times 75$ image as an input, and an array of track parameters as the output), with each sample containing a variable number of tracks described by a Poisson distribution ($N_{max} = 40, N_{mean} = 35$). 
More data (larger image for better granularity) is needed to reconstruct more than 40 tracks, which requires a larger neural network. The objective of this approach is to demonstrate that neural networks can be used for track reconstruction. It should be mentioned that the inspiration for this approach comes from a study carried out by \emph{M. Kucharczyk et al} \cite{91} which is capable of reconstructing only three tracks. Furthermore, their approach operates on $25 \times 25$ pixels (where every z-pixel corresponds to a `module'). 

\subsubsection{Results}
\begin{figure}[H]
\hbox{\hspace{+0.78em}\includegraphics*[width=1\linewidth]{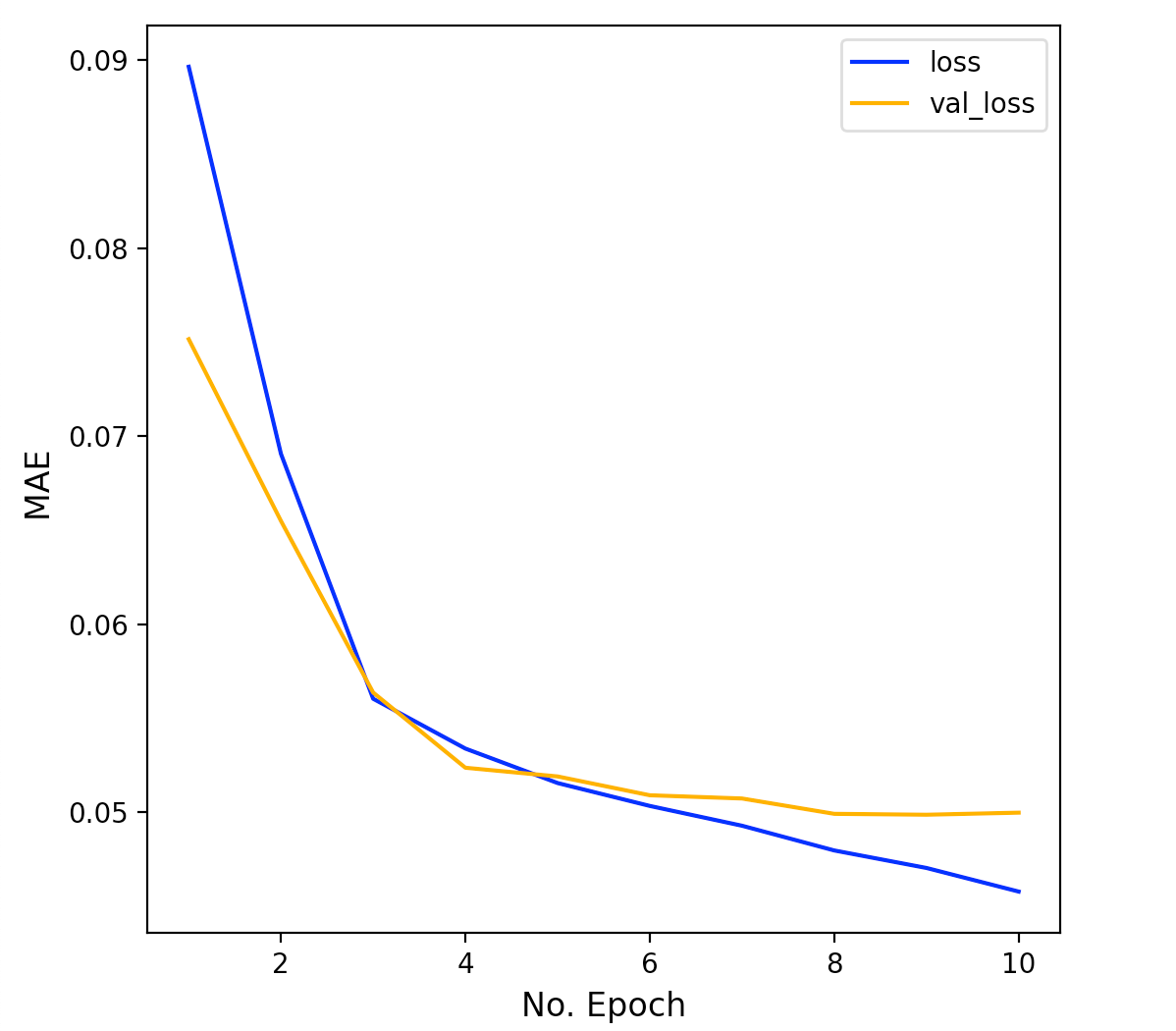}}
\caption{Mean average error (loss) vs number of epochs.}\label{fig:27}
\end{figure}

\begin{figure}[h]
\includegraphics*[width=0.9\linewidth]{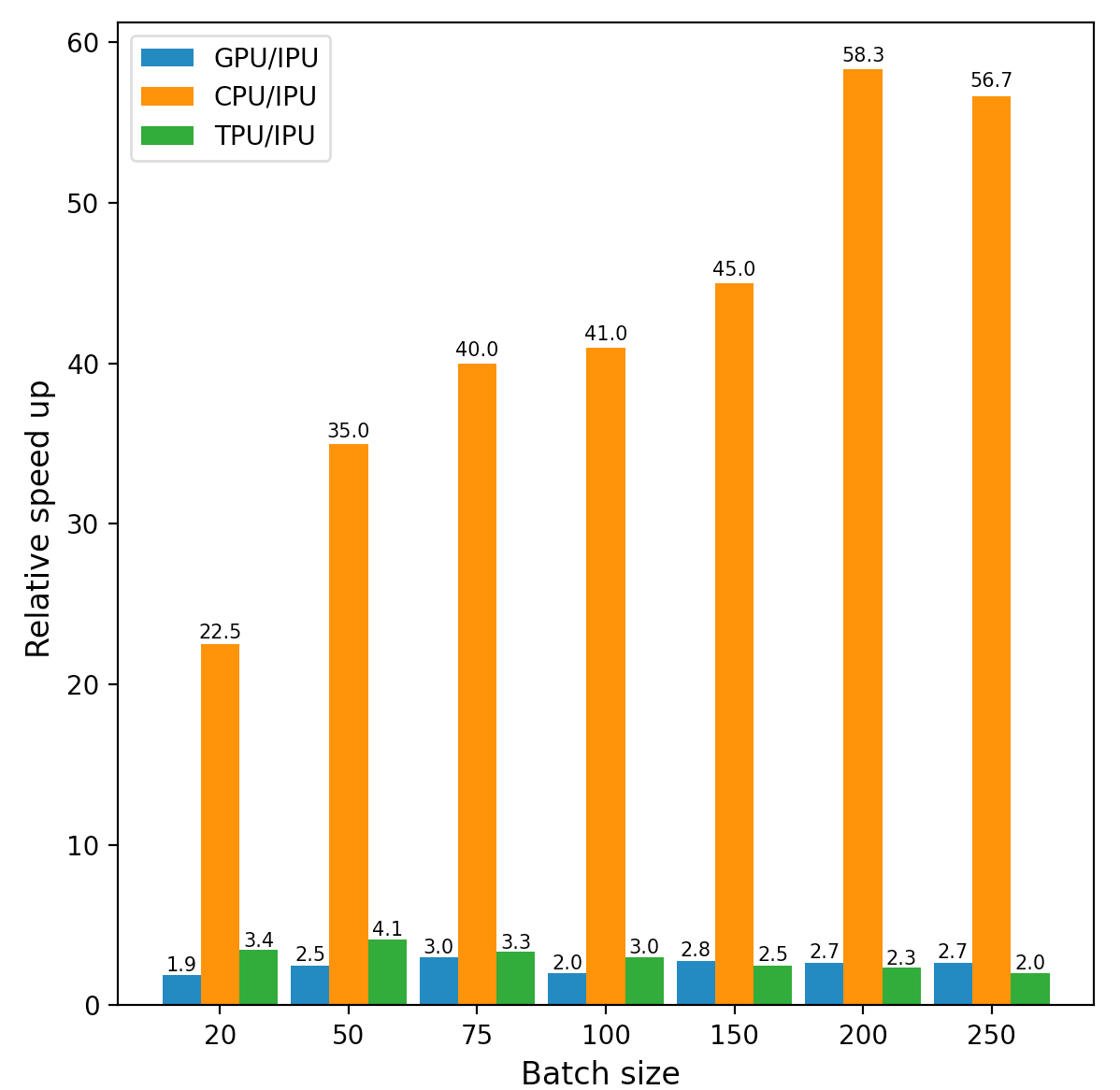}
\caption{Comparison of the time to train the IPU relative to the CPU, GPU and the TPU.} \label{fig:28}
\end{figure}

\begin{figure}[h]

\includegraphics*[width=0.9\linewidth]{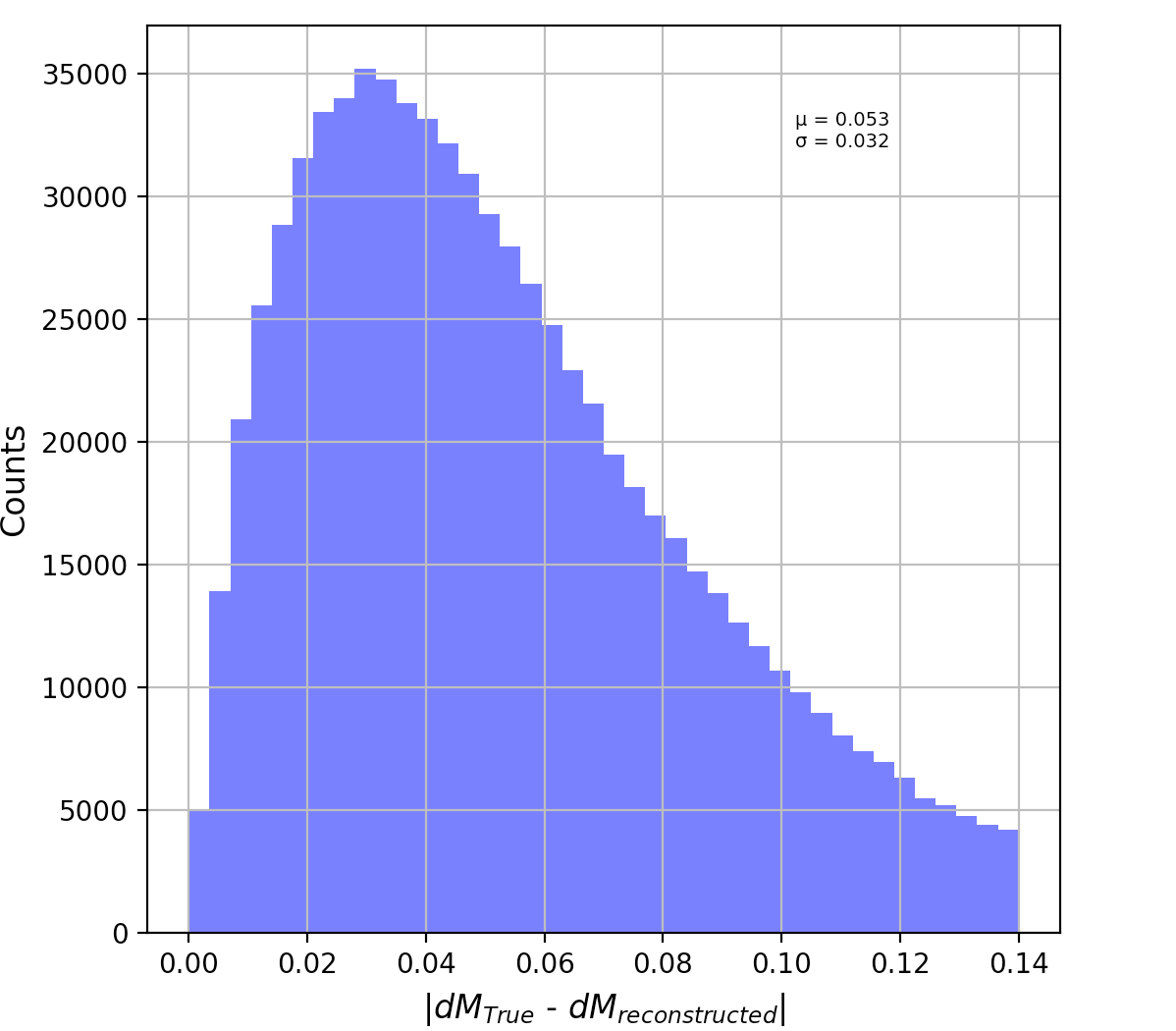}
\caption{Distribution of the absolute difference between the predicted and the true track parameters. } \label{fig:29}
\end{figure}

Table 4 summarises the reconstruction performance of the neural network as a function of the number of tracks. Figures \ref{fig:27}, \ref{fig:28}, and \ref{fig:29} respectively represent the MAE loss of the neural network as a function of the number of epochs, the training time (speed-up) of the neural network on different computing architectures, and the absolute difference between the predicted and target track parameters.

\begin{table}[h]
\begin{adjustbox}{width=\columnwidth,center}
\begin{tabular}{|c|c|c|c|c|c|}
\hline
\textit{Number of tracks} & \textit{5} & \textit{10} & \textit{20}                & \textit{30}                & \textit{40}               \\ \hline
\textit{Reconstruction efficiency}  & \textit{0.991 $\pm$ 0.005}  & \textit{0.975 $\pm$ 0.005}  & 0.939 $\pm$ 0.006 & \textit{0.899 $\pm$ 0.006} & \textit{0.854 $\pm$ 0.006} \\ \hline
\textit{Fake track fraction}      & \textit{0.047 $\pm$ 0.008}  & \textit{0.042 $\pm$ 0.005}  & \textit{0.046 $\pm$ 0.003}             & \textit{0.041 $\pm$ 0.004}             & \textit{0.031 $\pm$ 0.003}             \\ \hline
\textit{Clone track fraction}     & \textit{0.100 $\pm$ 0.017}  & \textit{0.118 $\pm$ 0.009}  & \textit{0.115 $\pm$ 0.006}             & \textit{0.122 $\pm$ 0.005}             & \textit{0.120 $\pm$ 0.005}             \\ \hline
\end{tabular}
\end{adjustbox}

\caption{Track reconstruction performance of the predictive combinatorial seeding algorithm as a function of the number of tracks.}
\label{tab:my-table}
\end{table}

The reconstruction efficiency (table 4) decreases with the number of tracks (a trend observed across all reconstruction algorithms discussed earlier). At about 40 tracks, both the combinatorial and the artificial retina algorithm are more efficient than the NN model. This is primarily due to the limited granularity of the input data the NN operates on (5625 pixels in total), suggesting that the performance can be improved if a larger image is treated on a more extensive network. The fake and clone track fractions increase with the number of tracks for the same reason as discussed earlier. Compared to the artificial retina algorithm, the NN model generates almost a quarter as many fake and clone tracks. This is due to the network design, which produces a fixed-sized output (if $N$ out of 40 tracks are observed, then the remaining $N-40$ parameters are padded to maintain a fixed output).

The NN was trained over ten epochs in batches of 250 training samples. Figure \ref{fig:27} shows the mean average error (loss) as a function of the number of epochs. The NN starts to overfit around the 6th epoch; this happens when a model performs well on the training data but not on the validation data. This is precisely due to the NN learning the noise patterns present in the training data instead of learning the true dependencies between the inputs and the outputs \cite{73}. 
Overfitting can be avoided by increasing the sample size and using dropout layers (turning off a random fraction of neurons).

This model scales independently with the number of Velo-hits/tracks since the whole Velo-space (regardless of the number of hits) is transformed into a $75 \times 75$ image. For this reason, the time to train the neural network is measured to gauge the performance of various computing architectures.
Training a NN requires an ample amount of linear algebra operations, as shown in eq(23). These operations are dense and regular (structured), and for this reason, they map efficiently on parallel multi-core architectures such as GPUs and IPUs. In terms of training time (figure \ref{fig:28}), the IPU on average is about 42x faster than the CPU and about 2.5x and 2.9x faster than the GPU and the TPU respectively. The performance of the GPU and the TPU is as expected (similar to the IPU) since both architectures are deeply optimised for ML applications. 
The number of training steps within a single epoch is defined as $N_{sample}/N_{batch size}$.
It is worth mentioning that there exists a delicate relation between the batch size and the accuracy of a neural network. Very large batch sizes would speed up the training process at the expense of the network's quality (accuracy). This is because large batches tend to converge to sharp minima of the loss function, resulting in a poorer generalization of the model \cite{74}. It should be noted that the IPU was unable to train this particular model in batches larger than 250, primarily due to the memory constraints of the IPU (286 MiB of total memory); in contrast, the CPU, GPU, and the TPU had access to far larger shared memories (\textbf{i.e} 396000, 16000, 64000 MiB respectively).

\subsection{The trackless vertex finder}
This novel approach proposes reconstructing the primary vertices before the track reconstruction stage, potentially simplifying the pattern recognition stage as a byproduct. 
\subsubsection{Design}
\begin{figure}[h]
\includegraphics*[width=0.86\linewidth]{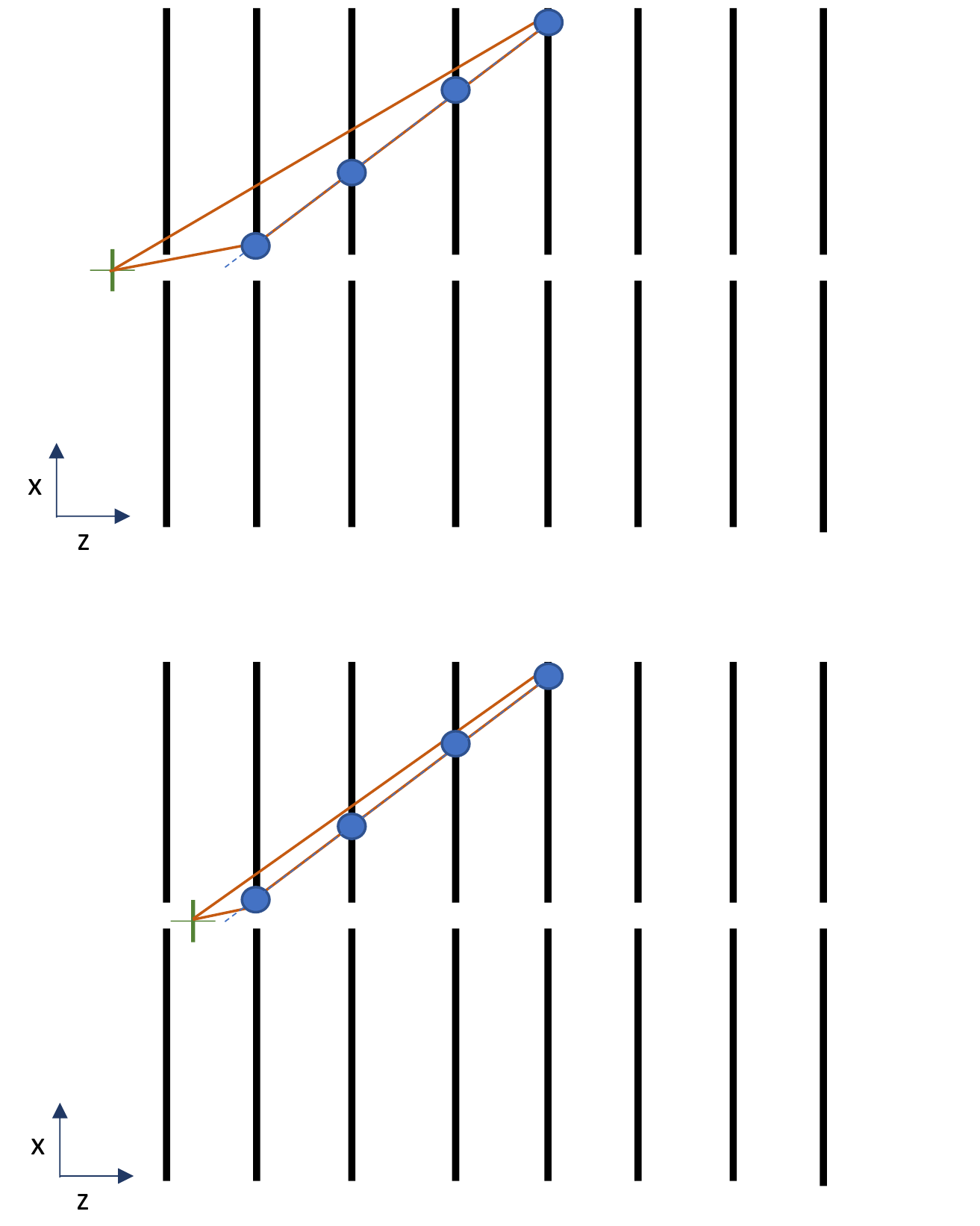}
\caption{Depiction of two polygons calculated by the trackless vertex finder. The image contains two separate sets of Velo-modules. The blue dots represent the hits, the green `plus' sign is the guess vertex, and the solid brown lines are the polygon's edges. An inaccurate guess generates a large polygon (top half), and conversely an accurate guess generates a small polygon (bottom half).} \label{fig:30}
\end{figure}

Given that all primary tracks (in a single event) emerge from a single vertex placed along the beamline, this algorithm estimates the primary vertex using a principle similar to that of the artificial retina algorithm.
The idea involves generating a 1D bin of z-coordinates (potential vertices); at each bin, the gradient ($arctan \ \varphi$) of all hits with respect to the guess vertex is calculated. Hits in every Velo-module are sorted in ascending values of $arctan \ \varphi$. For every sorted `row' of hits in subsequent modules (similar to the $\varphi_{0}$ window in figure \ref{fig:9}), the area ($A$) of a 5-sided polygon (the guess vertex connected with 4 adjacent hits) is calculated. This area is a measure of how well the guess vertex fits the data. A better guess corresponds to a smaller area. Conversely, an inaccurate guess corresponds to a larger area (figure \ref{fig:30} represents this idea). The following equation (analogous to the retina algorithm) determines the intensity of each guess vertex.
\begin{equation}
I_{j} = \sum_{n}^{} exp(-\frac{\sum_{i}A_{i}}{2\sigma })	
\end{equation}
Where $I_{j}$ is the intensity of the $j^{th}$ guess vertex, $i$ is the polygon index (belonging to a track), $n$ is the `row' index of the sorted hits, and sigma is a parameter that can be tuned to improve the sensitivity. Once the intensities of all guess vertices are computed, a 1D heatmap (intensity vs z position) is generated, and the peak here corresponds to the PV candidate.

\vspace{+3mm}
\vspace{-2mm}

\subsubsection{Results}
\begin{figure}[h]
\includegraphics*[width=0.87\linewidth]{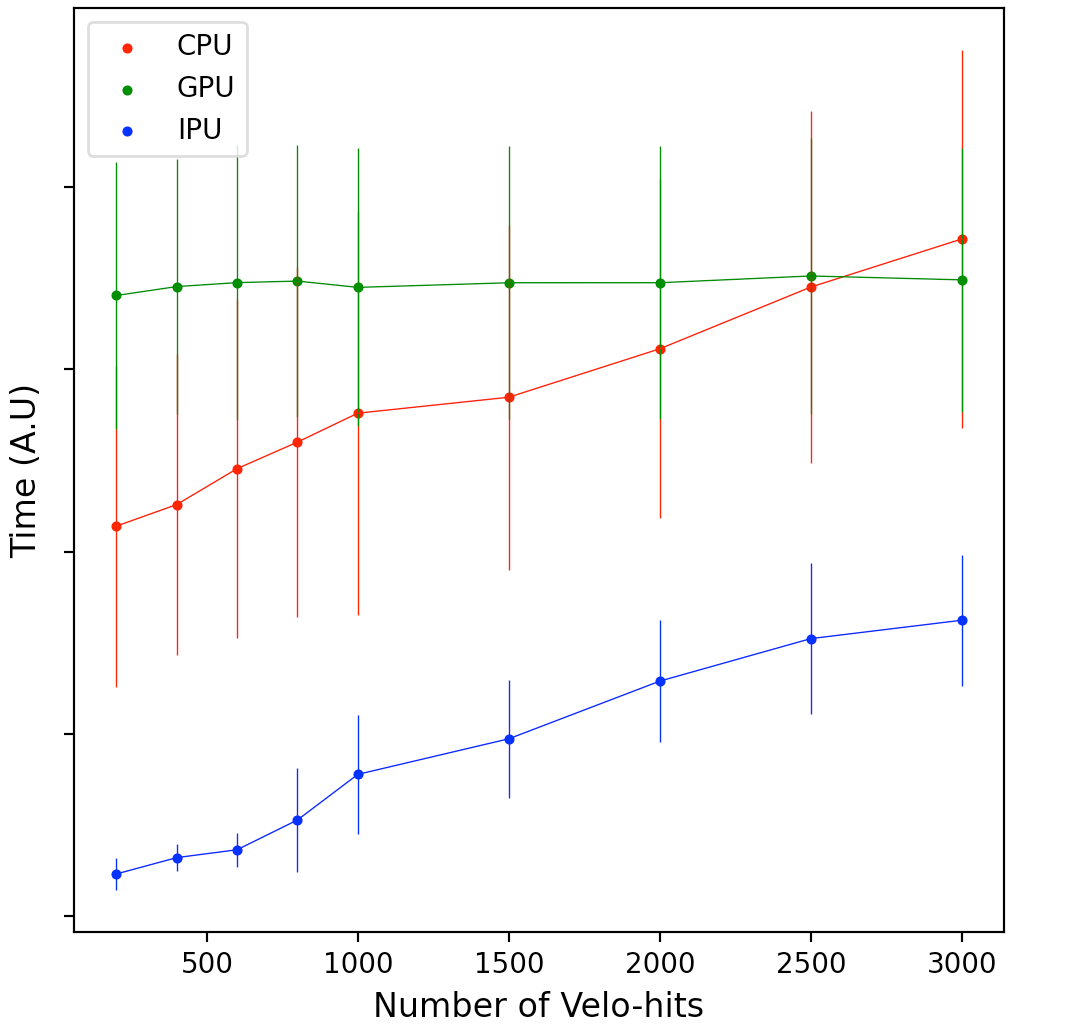}
\caption{Scaling of the computing time with the number of tracks, evaluated on the CPU, GPU, and the IPU.} \label{fig:31}
\end{figure}

\begin{figure}[h]
\includegraphics*[width=0.95\linewidth]{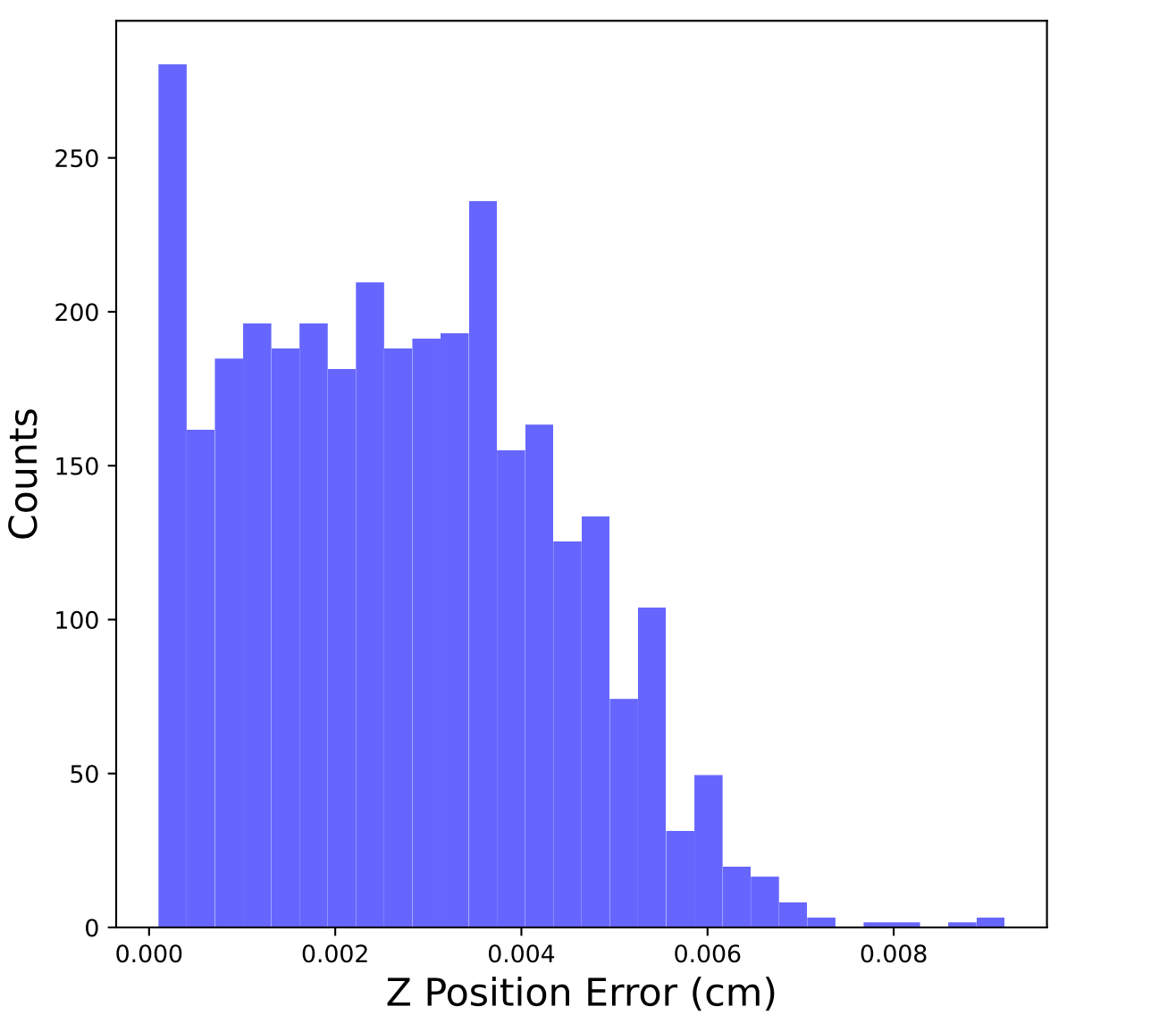}
\caption{Distribution of the absolute difference between the predicted and the true primary vertices generated using 2000 events, each containing about 200 reconstructable tracks (ms-data).} \label{fig:32}
\end{figure}

Figure \ref{fig:31} describes the computing time as a function of the number of Velo-hits on different computing architectures. Figure \ref{fig:32} depicts the absolute difference between the true and reconstructed vertices for 2000 individual events. 

The scaling of the computing time is similar to the retina algorithm (as expected). The IPU outperforms the GPU and the CPU. The speedup ratios (CPU/GPU and CPU/IPU) at 3000 hits are 1.1 and 1.6 respectively.
At smaller domains ($\leq$ 2500 hits), the CPU outperforms the GPU primary due to the overheads in decomposing the problem across the shared memory architecture of the GPU. The GPU scales constantly in the domain 1000 $\leq$ hits $\leq$ 32000 due to its sheer core count (hits greater than 3000 are not shown here due to how rapidly the CPU time scales relative to the  IPU and the GPU). The GPU, in-fact, outperforms the IPU at 30000 hits (roughly corresponding to 2700 Velo-tracks), mainly since the GPU contains roughly 2x cores and 56x memory relative to the IPU. 

\section{Discussion}
\subsection{Computing performance and limitations}
There are three main effects at play here that explain the large performance discrepancy (speed-up ratios: CPU/IPU, CPU/GPU) between ML and non-ML based algorithms.
\subsubsection{ML optimization}

TensorFlow is an open-source library designed specially from the ground up for Math and ML-based computations such as: designing various neural network architectures (DNNs, CNNs, RNNs, GANs) and training/inferencing such neural networks. The regular and dense yet relatively straightforward (linear algebra) nature of ML computations benefits more from a larger count of simpler cores (GPU, IPU) compared to a few complex cores (CPU). For this reason, large scale neural networks are almost always trained on multi-core architectures like the GPU and specially formulated hardware like the IPU and the TPU. As a consequence, ML-based TensorFlow operations (by design) are highly optimised on the GPU, IPU, and the TPU \cite{29,30,75}. 

\subsubsection{Parallelism}
The performance of non-ML based algorithms solely depends on how TensorFlow maps the XLA backend to the hardware. In comparison to parallel programming techniques such as OpenMP\footnote{Open Multi-Processing} and MPI\footnote{Message Passing Interface}, Tensorflow practically offers no control over the parallelization of the algorithm (\textbf{i.e} domain decomposition or functional decomposition), nor is it possible to gauge how non-ML algorithms are being parallelized on the CPU, GPU, and the IPU hardware. This, in fact, is the largest limitation of this study since it masks the true potential of parallel architectures.
A vital improvement to this research would involve rewriting the non-ML algorithms in the native parallel programming platforms of the GPU and the IPU, \textbf{i.e} CUDA and POPLAR respectively (both jointly support C++ at the moment). The limitation of TensorFlow's non-existent interface for fine-grained parallelism is particularly evident in the poor performance of the GPU whilst running the combinatorial seeding algorithm.

By using CUDA, the combinatorial algorithm can be modified such that each thread is assigned to a single hit within a module and is responsible for generating track segments from hits in the next module to the hit assigned to it (uniform workload). Once all track segments have been generated, a synchronization step is initiated, which uses the shared memory of the GPU to identify the best triplet track segments. This level of fine-grained parallelism would bypass the bottleneck (diverging workflows) the TensorFlow variant of the combinatorial algorithm faces, sharply improving the performance as a result. Examples from the literature support this argument: First, the GPU-based \emph{search by triplet} algorithm written in CUDA is 6.8 times faster than the CPU variant \cite{43}. Second, the CUDA-based \emph{cellular automaton} algorithm for the CMS experiment outperforms the baseline CPU setup by a factor of 64 \cite{39}. Third, the CUDA-based \emph{TPC Tracking} algorithm for the ALICE experiment offers a 50x speedup over the baseline CPU setup \cite{76}.
Finally, it is worth mentioning the efforts being made to develop MPI like interface for Tensorflow, these include Uber’s Horovod \cite{77} and Cray’s Machine Learning Plugin \cite{78}. 

\subsubsection{TensorFlow (frontend  vs backend)}
\begin{figure}[h]
\includegraphics*[width=1\linewidth]{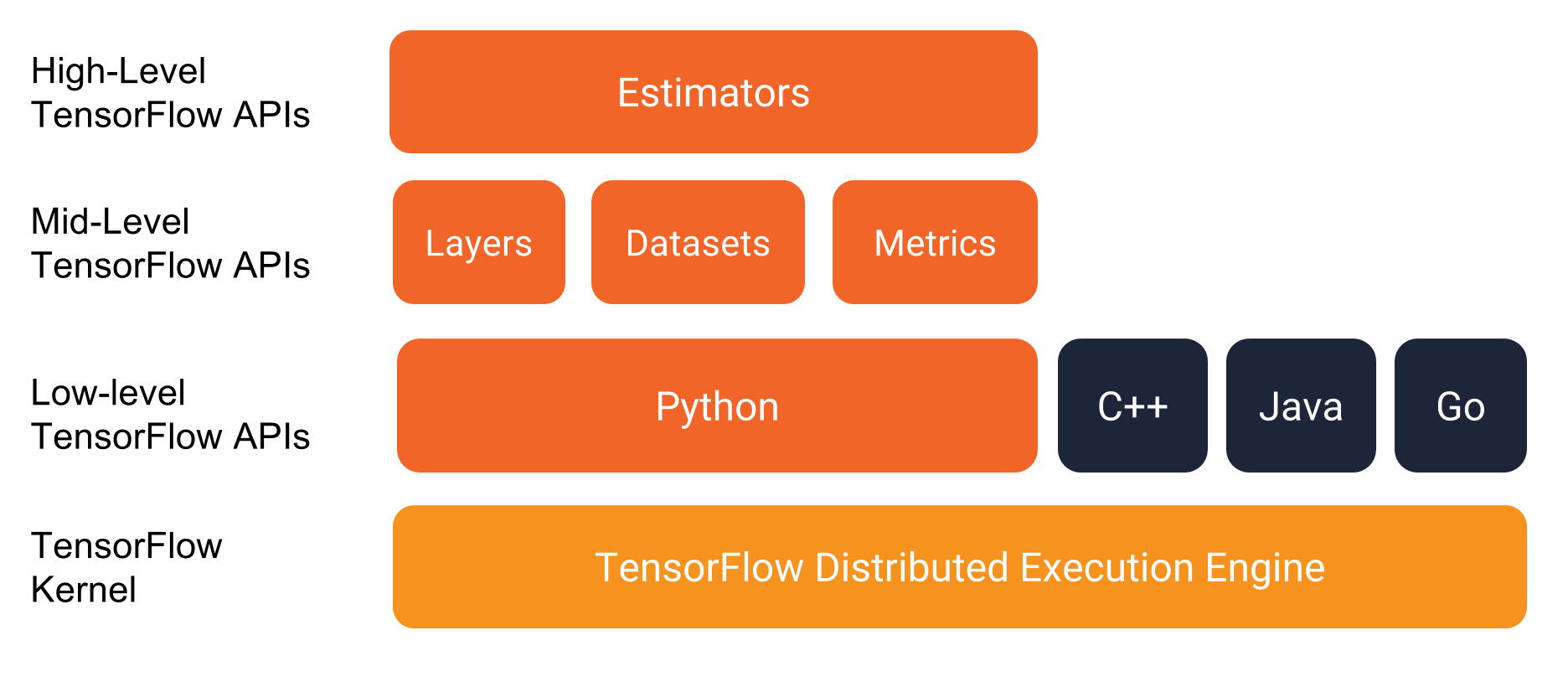}
\caption{TensorFlow's high-level representation. Figure Ref \cite{79}.} \label{fig:33}
\end{figure}

The core ML Tensorflow operations (backend kernel) are developed using Eigen (a high-performance C++ and CUDA numerical library). Tensorflow supports python, Go,  C++, and Java as frontend interfaces which are used to \emph{call} the core Eigen (C++) functions (figure \ref{fig:33} represents this hierarchy).

There are fundamental differences between C++ (ML backend) and pythonic Tensorflow (\textbf{i.e} combinatorial, artificial retina, and the trackless vertex finder algorithm). First, C++ is a compiled language that is directly converted into machine code (high-speed execution). Python, on the other hand, is an interpreted language that, when executed, is compiled into byte code by an interpreter, which is native instruction code for the Python virtual machine (slower execution) \cite{80}. Second, C++ is a statically typed language that requires objects/variables to have a predefined datatype (saves a significant fraction of time when using loops). In contrast, python dynamically assigns appropriate datatype for every variable (slowing the workflow) \cite{81}. To support this argument, consider the example of Numba (a Just-in-Time compiler that translates a subset of python code directly into machine code, analogous to TensorFlow in this context) \cite{82}. A recent study by \emph{L. Oden} compares the performance of GPU algorithms written in C-CUDA and Python-Numba. Results from \emph{Oden}'s study indicate that Numba's performance is comparable to C-CUDA; however, the part of the code that does not use the compiled Numba functions is interpreted as regular python code (unlike compiled C-CUDA), drastically degrading the performance of the algorithm \cite{83}. 

\vspace{-1mm}
\subsection{Other computing avenues in HEP}
This study involves benchmarking various computing architectures in the context of track reconstruction, an integral part of the trigger system, which by definition is executed in real-time (online). Further work is needed to assess the feasibility of such computing architectures; this would involve investigating other resource-intensive computing domains within HEP. Namely, the (offline) MC simulations, which consume about 70\% of the computing resources. The increase in efficiency and throughput of the upgraded HLT would require considerably more events to be simulated, requiring an increase of CPU resources by two orders of magnitude \cite{84}. The run 3 of the LHCb plans to reduce CPU resources spent on simulations by splitting them up into 40\% of full simulation, 40\% of fast simulations (lower accuracy faster processing), and 20\% of parametric simulations, which are described in detail in Ref \cite{85}.

First, full event simulations are challenging to develop for the GPUs since MC techniques by definition depend on branches that depend on random variables (\textbf{i.e} highly inefficient since threads would undergo diverging workflows). However, it is worth mentioning that GPUs can accelerate certain (dense, regular) tasks within MC simulations, \textbf{e.g}: the SIMD VecGeom library within the Geant4 simulation framework \cite{86}. Furthermore, it would be worth investigating the IPU's performance (relative to the CPU) in simulation workflows since the IPU is immune to performance penalties induced by diverging workflows (offering a lot more flexibility). 

Second, the need for faster simulations has yielded the use of GANs (Generative adversarial network) to generate events. GANs are a type of neural network architecture, which in the context of HEP, generate events by being trained on the full (slow) simulations. Once trained, GANs can generate parameterised outputs (generating specific events), offering a significant boost in the event-simulation rate compared to traditional methods \cite{87}. Due to their inherent ML nature, GANs map extremely well on ML optimised hardware. A recent study by \emph{L.R.M. Mohan et al} investigated the performance of two such networks (DijetGAN \& LAGAN) in the context of event generation \cite{88}. The event generation rate ratios (IPU/CPU, GPU/CPU) were found to be: (36,6) and (86,11) for the DijetGAN and LAGAN networks respectively, suggesting that multi-core architectures hold the potential to alleviate the computational burden in the regime of event simulation.

Additionally, ML-optimised hardware in HEP is of increasing importance since many domains within HEP are using/can benefit from ML approaches. In terms of event classification, the CMS collaboration uses a boosted decision tree (BDT) classifier to extract signatures of the Higgs decaying into a pair of muons \cite{89}. Moreover, neural networks are actively used for particle identification in the LHCb experiment. For example, \emph{ProbNN} is a multi-class classifier designed to distinguish between six charged particle classes: electron, muon, pion, kaon, proton, and software generated fake tracks \cite{90}.

\section{Summary and Conclusions}

This paper gives an overview of different computing techniques used in the LHCb experiment. Track reconstruction is the most computationally intensive task within HLT1. For this reason, the performance of three different Velo-track reconstruction algorithms and a vertexing algorithm were benchmarked on the CPU, GPU and a new type of processor known as the IPU. In all cases, the IPU outperforms the CPU and the GPU. In contrast, the CPU outperforms the GPU in one case (partly because the SIMD architecture of the GPU faces significant performance penalties due to diverging workflows). The computing performance (speed-up ratios: CPU/IPU, CPU/GPU) for ML and non-ML algorithms is of the order $\mathcal{O}(10)$ and $\mathcal{O}(1)$ respectively. This is mainly because of two reasons. First, ML-based computations (by design) are highly optimised to take advantage of $\mathcal{O}(1000)$ cores of the IPU and the GPU. Second, the non-ML algorithms were written using TensorFlow, which offers no control over parallelism, resulting in poorly optimised workflows. Despite these severe limitations, the multi-core architectures outperform the CPU (in 4/4 and 3/4 cases on the IPU and GPU respectively). These results are encouraging given the fact that a four-fold shortage in computing power is forecasted by 2030. A vital improvement to this line of research would involve rewriting the non-ML algorithms in the native programming platforms of the GPU and the IPU. Moreover, additional work is needed to explore how well such architectures map on to other computing domains within HEP, mainly, event selection, particle identification and the resource-intensive MC simulations.

\balancecolsandclearpage

\end{document}